\documentclass[fleqn,usenatbib]{mnras}
\usepackage{bm}
\usepackage{newtxtext,newtxmath}
\usepackage{comment}

\usepackage[T1]{fontenc}
\usepackage{xcolor}
\usepackage{subcaption}
\usepackage[
    starfontserif 
    ]{starfont}

\DeclareRobustCommand{\VAN}[3]{#2}
\let\VANthebibliography\thebibliography
\def\thebibliography{\DeclareRobustCommand{\VAN}[3]{##3}\VANthebibliography}

\usepackage{graphicx}	
\usepackage{float}
\usepackage{amsmath}	
\graphicspath{{Figures/}}
\usepackage{booktabs} 
\usepackage{multirow}

\title[Most Gas Dwarfs are Molten]{Most Rocky Sub-Neptunes are Molten: Mapping the Solidification Shoreline for Gas Dwarf Exoplanets}

\author[Robb Calder et. al.]{
Robb Calder,$^{1}$\thanks{E-mail: rdc49@cam.ac.uk}
Oliver Shorttle,$^{1,2}$
Harrison Nicholls,$^{1,3}$
Tim Lichtenberg,$^{4}$
Claire Marie Guimond$^{3,5}$
\\
$^{1}$Institute of Astronomy, University of Cambridge, Madingley Road, Cambridge CB3 0HA, UK\\
$^{2}$Department of Earth Sciences, University of Cambridge, Downing Street, Cambridge CB2 3EQ, UK\\
$^{3}$Atmospheric, Oceanic, and Planetary Physics, Department of Physics, University of Oxford, Oxford OX1 3PU, United Kingdom\\
$^{4}$Kapteyn Astronomical Institute, University of Groningen, P.O. Box 800, 9700 AV Groningen, The Netherlands\\
$^{5}$Department of Earth and Planetary Sciences, ETH Zurich, Sonneggstrasse 5, 8092 Zurich, Switzerland\\
}

\date{Accepted XXX. Received XXX; in original form XXX}

\pubyear{2025}

\begin{document}
\label{firstpage}
\pagerange{\pageref{firstpage}--\pageref{lastpage}}
\maketitle

\begin{abstract}
Sub-Neptunes are the most common type of detected exoplanet, yet their observed masses and radii are degenerate with several interior structures. One possibility is that sub-Neptunes have silicate/iron interiors and H$_2$-dominated atmospheres \textcolor{black}{($\mu$<3.8\,g mol$^{-1}$)}, i.e., they are `gas dwarfs'. If gas dwarfs have molten interiors, interactions between their magma oceans and atmospheres will produce distinct observational signatures. These signatures may break the degeneracy in interior structure, while providing insight into their interior processes, history, and population trends. We expect all such planets are born molten, but under what conditions do they remain molten today? We use the coupled interior-climate evolution model, \texttt{PROTEUS}, to estimate the `solidification shoreline': the instellation flux boundary (as a function of stellar $T_{\rm eff}$) that separates molten gas dwarfs from solidified ones. Our results show that 98\% of detected sub-Neptunes occupy a region of parameter space consistent with their having permanent magma oceans, if they are gas dwarfs. While mantle $f{\rm O}_2$ and bulk volatile C/H ratio both influence magma ocean \textcolor{black}{cooling}, planets with oxidising mantles and carbon-rich atmospheres \textcolor{black}{are likely to have high mean-molecular weight atmospheres ($\mu$>3.8\,g mol$^{-1}$) and are thus outside the scope of this study}. Therefore, most detected sub-Neptunes, if they are gas dwarfs, have permanent magma oceans. This result motivates further research into the interactions between molten interiors and overlying atmospheres, and campaigns to identify unambiguous signatures of these interactions.
\end{abstract}


\section{Introduction}
\label{sec:introduction}

Exoplanets in the sub-Neptune regime \citep[$1.8\,R_{\oplus}\!<\!R\!<\!4\,R_{\oplus}$;][]{Fulton2017} are at the forefront of exoplanet research, with no analogue in our Solar System. The Kepler survey has revealed that they are the most common type of discovered exoplanet \citep{Howard2012,Thompson2018,Hsu2019}, so their study is essential for a unified theory of planetary formation and evolution. With the launch of the JWST and the upcoming launch of Ariel \citep{Changeat2025}, detailed atmospheric characterisation of sub-Neptunes via transmission spectroscopy is now possible. Recent JWST observations of sub-Neptunes have detected molecules such as H$_2$O, CO$_2$ and and CH$_4$ at confidence levels ranging from 3$\sigma$ to 5$\sigma$ \citep{Madhusudhan2023,Benneke2024,Holmberg2024,Hu2025}. Furthermore, the upcoming PLATO mission is expected to detect several hundred sub-Neptunes that are ideal for follow up observations with JWST and Ariel \citep{Rauer2025}. Therefore, the study of sub-Neptunes is one of the most important and timely areas of research in planetary science.

One of the key challenges in the study of sub-Neptunes is unveiling the nature of their interior structures, given that their observed masses and radii are consistent with both volatile-rich and volatile-poor scenarios on a population level \citep{Howe2014,Rogers2023,Parc2024}. Two possible hypotheses for their interior structures, among others, are the `Gas Dwarf' and `Water World' hypotheses. Gas dwarfs are expected to have iron-silicate interiors and H$_2$-dominated envelopes \textcolor{black}{($\mu$<3.8\,g mol$^{-1}$)} constituting 0.1–10\% of the total planetary mass \citep{Wolfgang2015,Ginzburg2016,Tang2024}, and water worlds are expected to have H$_2$O mass-fractions of 10-50\% \citep{Zeng2019,Venturini2020,Lee2021,luque_Density_2022,Burn2024} \textcolor{black}{(see table \ref{tab:gasdwarfwaterworldcriteria} and section \ref{subsec:gasdwarfwaterworlddefinitions} for our explicit definitions of the terms `gas dwarf' and `water world')}. The degeneracy between gas dwarfs and water worlds is made more pressing by the fact that a hypothesised subset of the water world population, `hycean worlds' \citep{Madhusudhan2021,Rigby2024_b}, have been suggested to have habitable surface conditions. This possibility has been argued with claims of potential biosignature detections in the atmosphere of a candidate hycean world \citep{Madhusudhan2023,Madhusudhan2025,Pica-Ciamarra2025}. Therefore, a framework for constraining the thermal, compositional and structural make-up of sub-Neptunes is essential not only for a comprehensive understanding of planet formation and evolution, but also for refining the bounds on where habitable conditions may be found in the universe \citep{Lichtenberg2025}.

One method for discriminating between gas dwarfs and water worlds is through observations of their upper-atmospheric chemistry via transmission spectroscopy. This is done by searching for observational signatures of specific molecules in the atmospheres of sub-Neptunes \citep{Constantinou2022,Madhusudhan2023,Benneke2024}, and comparing these observations with models that relate their upper atmospheric chemistry to their composition, structure and climate state \citep{Shorttle2024,Cooke2024,Wogan2024,Rigby2024}. For example, it has been suggested that a lack of NH$_3$, alongside a detection of CO$_2$ and CH$_4$, in a sub-Neptune atmosphere is evidence of the presence of a liquid water ocean \citep{Madhusudhan2021,Madhusudhan2023,Hu2025}, the argument being that the dissolution of NH$_3$ in a liquid-water ocean is the only explanation for its absence alongside CO$_2$ and CH$_4$. However, it has also been suggested that if a gas dwarf were to instead have a magma ocean in contact with its atmosphere, this would also act as a solvent for N-bearing species, complicating the use of an NH$_3$ non-detection alone to discriminate between the hycean worlds and gas dwarfs \citep{Shorttle2024,Glein2025}. This means that there is an observational degeneracy between hycean worlds and gas dwarfs with magma oceans.

As well as being observationally degenerate with other structural scenarios for sub-Neptunes, if gas dwarfs were to have surface magma oceans, these magma oceans could offer valuable insights into the interior composition and population statistics of sub-Neptunes. For example, if an H$_2$-dominated atmosphere is in contact with a magma ocean, then it will interact chemically with volatiles dissolved in the magma \citep{Schaefer2017,Kite2019,Tian2024,Horn2025,Miozzi2025}. Studies have shown that the resulting atmospheric composition will be strongly affected by the oxidation state as well as the volatile content of the magma ocean, potentially allowing us to constrain the interior redox conditions of gas dwarfs via transmission spectroscopy \citep{Shorttle2024,Rigby2024,Bower2025} and emission spectroscopy \citep{Nicholls2025}. Also, the underabundance of sub-Neptunes with $R\!>\!3\,R_{\oplus}$ could be a consequence of the reduction of sub-Neptune radii due to the dissolution of H$_2$ in magma oceans \citep{Kite2020}. Given the implications magma oceans have for the climate, habitability and atmospheric compositions of gas dwarfs, it is pertinent to ask: how common are magma oceans on gas dwarfs? While all gas dwarfs are expected to have magma oceans at the point of formation, due to the heat from accretion, a subset of the population may experience surface heating rates sufficient to remain molten permanently, rather than solidifying. Therefore, the key issue is identifying the fraction of the gas dwarf population that have permanent magma oceans.

Determining the fraction of gas dwarfs with molten surfaces requires accurate modelling of their climate and the heat transport within their silicate interiors. Given that gas dwarfs have atmospheric mass fractions much larger than those expected for Earth-sized planets, the greenhouse effect in these atmospheres is expected to produce surface temperatures far exceeding those of Earth-sized planets. This would lead to molten surfaces at instellation fluxes lower than those required to induce molten surfaces on Earth-sized planets \citep{Lichtenberg2021,Innes2023,Rigby2024,Tang2024}. Volatile exchange between the magma and the atmosphere sets the chemical composition of the atmosphere \citep{Schaefer2017,Kite2019,Tian2024,Shorttle2024,Rigby2024}, which affects the greenhouse heating of the surface \citep{Lichtenberg2021,Boer2025}, which in turn affects the surface melt-state. Moreover, even if sub-Neptunes start out in a molten state and solidify over time, this cooling is expected to occur over gigayear timescales \citep{Vazan2018,Tang2024}. The magnitude of these cooling timescales implies that many sub-Neptunes may be young enough \citep{sandoval_TheInflu_2021, Bean2021} such that they still have magma oceans today, even if they will eventually solidify. The variety of processes that affect the cooling of magma oceans suggests that the question of whether gas dwarfs will remain permanently molten will vary considerably across the population, implying that answering this question on a population level is challenging.

However, while there are many complex physical and chemical processes that can influence the cooling of magma oceans on planets in general, there are constraints imposed by the gas dwarf hypothesis that limit the extent to which these processes can vary across this sub-population. For example, the gas dwarf hypothesis requires that the planetary atmosphere is H$_2$-dominated, which limits the extent to which the atmospheric composition can vary, and thus the effect of composition on surface heating. For this reason, it is reasonable to assume that, to first order, the question of whether gas dwarfs as a population will remain permanently molten will be determined by their instellation flux and envelope mass fraction. These limitations imposed by the gas dwarf hypothesis allow us to define a `solidification shoreline' that divides this parameter space into a region where gas dwarfs are expected to solidify and a region where they are expected to remain permanently molten, similar to concepts such as the habitable zone \citep{Hunag1960,Kasting1993,Kopparapu2013,Turbet2023} and the cosmic shoreline \citep{Zahnle2017,Lustig2019,Moran2023,Chatterjee2024}. Nevertheless, it is still necessary to explore the effect of atmospheric composition, and other parameters, on the location of this shoreline. This requires coupling of the evolution of the rocky interior, the climate, and atmospheric chemistry of sub-Neptunes.

Previous studies modelling the behaviour of magma oceans on gas dwarfs have either lacked the required coupling between the interior and the atmosphere, or have not considered their time-dependent evolution \citep{Huang2022,Shorttle2024,Rigby2024,Breza2025,Nixon2025}. Progress has been made to model the thermal evolution of magma oceans on super-Earths and sub-Neptunes \citep{Vazan2018,Herath2024}, and some of these models couple the evolution of the rocky interior and the atmosphere \citep{Lehmer2017,Kubyshkina2020,Tang2024}. However, none of these models account for volatile exchange between atmospheres and deep magma oceans, nor do they  estimate what fraction of the sub-Neptune population has surface magma oceans. \cite{Breza2025} estimated the fraction of sub-Neptunes that may host surface magma oceans using a static interior structure model. However, as stated in their work, a thorough investigation of this research question requires an evolutionary modelling framework.

In this work, we apply a 1D coupled interior-climate evolution model \citep{Lichtenberg2021,Nicholls2024,Nicholls2025,lichtenberg_proteus_2026} to simulate the thermal, compositional and structural evolution of sub-Neptunes, to determine when a permanent magma ocean is a viable steady state outcome. We consider a parameter space composed of stellar effective temperature, planet instellation flux, mantle oxygen fugacity, volatile C/H ratio and planet mass. In section \ref{sec:methods}, we discuss the modelling framework, \texttt{PROTEUS}, and the empirical scaling relations used in this study. In section \ref{sec:solidshoreline}, we report the results of our main study, where we determine the instellation flux as a function of stellar effective temperature at which the thermal steady state of a gas dwarf transitions from a permanent magma ocean to mantle solidification; i.e., the `solidification shoreline'. We also discuss the evolutionary timescales resulting from the models in this study. In section \ref{sec:secondaryparameters}, we show the results of our secondary parameter studies, where we investigate the sensitivity of the location of the solidification shoreline to mantle oxygen fugacity, planet mass and the C/H ratio of the bulk volatile inventory. In section \ref{sec:discussion}, we discuss our results in the context of the population of observed sub-Neptunes and how the solidification shoreline might be developed as a metric. We present our conclusions in section \ref{sec:conclusions}.

\section{Methods}
\label{sec:methods}

\subsection{Model Description}
\label{subsec:modeldescription}

To simulate the thermal evolution of gas dwarfs, we use the \texttt{PROTEUS} modelling framework \citep{Lichtenberg2021,Nicholls2024,Nicholls2025}. \texttt{PROTEUS} self-consistently simulates the coupled thermal evolution of a planet's interior and atmosphere, assuming that the interior starts in a fully molten state. \texttt{PROTEUS} couples the 1D mantle dynamics model \texttt{SPIDER} with the 1D radiative-convective climate model \texttt{AGNI} \citep{Nicholls2025b}, while allowing for volatile exchange between the semi-molten interior and the atmosphere. Volatile partitioning is calculated with the \texttt{CALLIOPE} model, which uses equilibrium gas chemistry and empirical solubility laws to partition volatiles between the magma and the atmosphere \citep{Lichtenberg2021,Bower2022,Nicholls2025}. The simulated evolution terminates when either the planet solidifies or achieves global energetic equilibrium between stellar irradiation, internal heat production, and atmospheric thermal emission to space. This energetic steady state corresponds to the permanent magma ocean scenario. This modelling framework allows us to simulate the thermal evolution of a gas dwarf self-consistently, allowing for either a solidified mantle or global energetic equilibrium with a semi-molten interior as a thermal steady state.

\textcolor{black}{Given the large number of parameters in our model, we specify these parameters in table \ref{tab:constantparameters}, alongside their fiducial values and our justification for these fiducial valuess. Also, see appendix \ref{apsec:generalsensitivitytests} for our general sensitivity tests, where we investigate the effects of varying several of these parameters on the results generated by PROTEUS.}

\begin{table*}
    \centering
    \renewcommand{\arraystretch}{1.2}
    \begin{tabular}{|p{7cm}|p{2cm}|p{8cm}|}
        \hline
        Parameter Name & Fiducial Value & Justification for Fiducial Value \\
        \hline
        Stellar T$_{\text{eff}}$ & 4500\,K & Midpoint in our parameter space \\
        Stellar Instellation & 22\,F$_{\oplus}$ & Instellation corresponding to solidification shoreline at T$_{\text{eff}}$=4500\,K \\
        Stellar Zenith Angle & 48.19\,$^{\circ}$ & Zenith angle from \cite{Cronin2014}\\
        Planet Mass & 5\,M$_{\oplus}$ & Average sub-Neptune mass from \cite{Madhusudhan2021} \\
        Core Radius Fraction & 0.55 & Earth Reference Value \\
        Core Density & 10738\,kg m$^{-3}$ & Earth Reference Value  \\
        Core Heat Capacity & 880\,J K$^{-1}$ kg$^{-1}$ & Earth Reference Value  \\
        Boundary Layer Conductivity & 2\,W m K$^{-1}$ & Thermal Conductivity of Basaltic magma \citep{LESHER2015}\\
        Boundary Thickness & 1\,cm & Taken from \cite{solomatov2015} \\
        Number of Spectral Bands in Radiative Transfer Model & 48 & Balance between numerical performance and resolution \\
        Number of Levels in Climate Model & 40 & Balance between numerical performance and resolution \\
        Initial Top-of-Mantle Heat Flux & 10$^5$\,W m$^{-2}$ & Initialisation of heat flux at hot temperatures \\
        Initial Specific Entropy & 3000\,J kg$^{-1}$ K$^{-1}$ & Initialising the planet fully molten \\
        Rheological Transition Value  & 0.6 & Covers range of experimental values in \cite{Costa2009} \\
        Rheological Transition Width & 0.2 & Covers range of experimental values in \cite{Costa2009} \\
        Reference Solid Viscosity ($\eta_s$) & 10$^{21}$\,Pa s & Solid Viscosity from \cite{Abe1993} \\
        Reference Melt Viscosity ($\eta_m$) & 10$^{2}$\,Pa s & Melt Viscosity from \cite{Abe1993} \\
        Number of Levels in Interior Model & 100 & Balance between numerical performance and resolution \\
        Mantle fO$_2$ ($\Delta \text{IW}$) & -3 & Chosen to result in reducing atmosphere\\
        Mantle Abundance of K Radioistopes at 4.55\,Gyr & 310\,ppmw & Earth Reference Value \\
        Mantle Abundance of U Radioistopes at 4.55\,Gyr & 31\,ppb & Earth Reference Value \\
        Mantle Abundance of Th Radioistopes at 4.55\,Gyr & 124\,ppb & Earth Reference Value \\
        Volatile H Mass (fraction of mantle mass) & 709\,ppm & Required to achieve EMF of 0.1\,\% \\
        Volatile C/H Mass Ratio & 0.32 & 100$\times$ solar metallicity \\
        Volatile N/H Mass Ratio & 0.09 & 100$\times$ solar metallicity \\
        \hline
    \end{tabular}
    \caption{\textcolor{black}{The parameters of the \texttt{PROTEUS} model, their fiducial values in this study, and our justification for each fiducial value.}}
    \label{tab:constantparameters}
\end{table*}

\subsubsection{Mantle Evolution Module}
\label{subsubsec:mantleevolution}

SPIDER \citep{Bower2018,Bower2022} numerically evolves the 1D profiles of temperature, specific entropy, melt fraction, and the thermodynamic quantities of a rocky planet’s mantle as a function of time. It accounts for convective heat flux, conduction, phase separation, gravitational settling and radiogenic heat production. An aggregate viscosity of the melt and solid material is calculated using the reference melt and solid viscosities from \cite{Abe1993}:

\begin{equation}
\begin{gathered}
\text{log}_{10}(\eta)=z\text{log}_{10}(\eta_m)+(1-z)\text{log}_{10}(\eta_s)\\
z(\phi)=\frac{1}{2}\left(1+\text{tanh}\left(\frac{\phi-\phi_c}{\phi_w}\right)\right),
\end{gathered}
\label{equ:aggregateviscosity}
\end{equation} \textcolor{black}{where $\phi$ refers to the melt fraction, $\phi_c$ refers to the melt fraction corresponding to the rheological transition, $\phi_w$ refers to the rheological transition width, $\eta_m$ refers to the reference melt viscosity and $\eta_s$ refers to the reference solid viscosity. This choice of rheological transition width and value reflect the experimental uncertainties in the rheological transition for magma described in \cite{Costa2009}. This equation reflects the trend seen in the semi-empirical model of \cite{Costa2009}. See appendix \ref{apsec:generalsensitivitytests} for an investigation into the effects of viscosity uncertainties on our results.}

The thermal evolution of the core is not explicitly modelled; it is treated as a heat reservoir at the bottom of the mantle with a constant density of 10,738 kg m$^{-3}$ and a constant heat capacity of 880 J K$^{-1}$ kg$^{-1}$, corresponding to that of a preliminary reference earth model adjusted for compositional differences \citep{Bower2018}. The radius of the core-mantle system is calculated for a given mass, using an equation of state for pure-MgSiO3 \cite{Bower2018}. We do not account for the effect of dissolved volatiles on the mantle density. The core interior-radius fraction is set to a value equivalent to that of the Earth: 0.55 \citep{Ladders1998}. For the study in which we vary the planet mass (section 4.3), we set the core radius fraction to a value of 0.7 to avoid numerical issues with the solver. These numerical issues arise due to high pressures in the deep mantle, where the thermodynamics and equation of state are poorly constrained \citep{Bower2018}. The mantle is assumed to begin its evolution with an adiabatic pressure-temperature profile and a surface heat flux of 105 W m$^{-2}$. The initial entropy is set to 3000 J K$^{-1}$ kg$^{-1}$. Long-term evolution of small planets is relatively insensitive to the exact value of initial entropy provided the internal temperature is fully above the mantle liquidus \citep{Lichtenberg2021,Schaefer2017}. Orbital evolution and tidal heating are not included in our model. If strong tidal heating were expected in the sub-Neptune population, then our simulations would represent a lower limit on interior heat generation and therefore provide a conservative estimate of whether sub-Neptunes are likely to be molten \citep{Farhat2025ApJ,Nicholls2025d}.

\textcolor{black}{Previous theoretical studies assume the existence of a solid, conductive layer between the surface of a magma ocean and the atmosphere \citep{Lebrun2013,schubert2015,schaefer_PREDICTIO_2016}, similar to those that form on lava ponds on Earth, that acts to limit heat transport between the mantle and the atmosphere.} As in previous modelling, we implement a parametrised conductive boundary layer at the mantle-atmosphere interface \citep{elkins_Linkedma_2008, Nicholls2024}. The temperature at the topmost layer of the mantle is used to calculate the temperature at the bottom of the boundary at each time-step, assuming a boundary layer thickness of 1\,cm \textcolor{black}{(see appendix \ref{apsec:generalsensitivitytests} for sensitivity tests for the boundary layer thickness). We also assume a thermal conductivity of 2 W m$^{-1}$ K$^{-1}$ for this layer, which is the thermal conductivity of basaltic magma \citep{LESHER2015}. This value will not change significantly within the range of likely silicate melt compositions \cite{LESHER2015}. Non-silicate melt compositions are outside the scope of this study.}

\subsubsection{Volatile Outgassing Module}
\label{subsubsec:outgassingmodule}

The surface temperature and mantle melt fraction (by mass) are then used to determine the chemical composition of the atmosphere using the \texttt{CALLIOPE} equilibrium outgassing model \citep{Bower2022,Nicholls2024,Boer2025}. In this work, we use \texttt{CALLIOPE} to solve for the partial pressures of eight volatile species: H$_2$, H$_2$O, CH$_4$, CO, CO$_2$, O$_2$, NH$_3$ and N$_2$. Sulfur chemistry is not included due to its negligible impact on climate in an H$_2$-dominated atmosphere \textcolor{black}{(see appendix \ref{apsec:sulfurtests}). We do not include silicon chemistry because the radiative properties of major silicon-bearing species are poorly constrained \citep{Ito2025}, and appreciable silicon in the atmosphere would increase the mean molecular weight of the atmosphere beyond the gas-dwarf paradigm we explore (see Section \ref{subsec:gasdwarfwaterworlddefinitions}).} We determine the partial pressures of the volatile species using empirically-derived solubility laws  \citep{ONeil2002,Ardia2013,Armstrong2015,DASGUPTA2022,Gaillard2022,SOSSI2023} and equilibrium thermochemical coefficients \citep{Stull1965,Chase1982} (see appendix A for the details of the solubility laws). 

We specify the following parameters as input to the calculation of the composition of the outgassed atmosphere: the total mass of hydrogen in the mantle-atmosphere system (as fraction of the mantle mass), the C/H and N/H ratios of the bulk volatile inventory, and the oxygen fugacity ($f_{{\rm O}_2}$) of the near-surface mantle. The oxygen fugacity is quantified in our model in log-units relative to the $f_{{\rm O}_2}$ of the iron-w\"ustite buffer, which is assumed to only vary with temperature in our model. The oxygen fugacity is calculated using equation 23 and the activity coefficient from table 3 in \cite{ONeil2002}. This activity coefficient was measured by \cite{ONeil1993} for temperatures ranging from \textcolor{black}{700\,K to 1433\,K at a reference pressure of 10$^5$\,Pa (1 bar). Given that surface pressures reach as high as 9\,GPa in our simulations, the adopted parametrisation of the iron-wu\"stite buffer may under-estimate the oxygen fugacity by up to 3 units in log-space \citep{Fischer2011}. This is comparable with empirical uncertainties in the redox-buffer parametrisations themselves. However, our modelling is agnostic to specific values of fO$_2$ since we do not make specific connections with iron distribution profiles in the mantle. Through our constraints on studied mean molecular weights, and thus atmospheric compositions, our results are insensitive to uncertainties in redox-buffer parametrisations and their pressure-dependencies.}

\textcolor{black}{
Several of the solubility laws and chemical equilibrium calculations in our model do not account for fugacity, i.e., they assume that the ideal gas law holds. However, given the range of total volatile inventory masses investigated in this study, we do not expect surface pressures and atmospheric compositions to deviate from the ideal gas assumption to the extent that our results are affected significantly. The maximum total volatile inventory mass in our study, which results in an EMF of \textasciitilde1\,\%, is 1\,\% of the mantle mass. \cite{Bower2025} show that, for total volatile inventory masses up to this value, the surface pressures that result from outgassing for a sub-Neptune at 3000\,K only deviate from the ideal gas scenario by a factor of up to 1.5. Furthermore, they also show that, for the same range of volatile inventories, the partial pressure of water in the atmosphere differs by less than a factor of 1.5 between the real-gas outgassing calculations and the ideal-gas outgassing calculations (carbon solubility in their work does not change appreciably as a result of fugacity corrections).}

\textcolor{black}{One caveat to the above point is that we go on to extrapolate our results, when comparing to the observed exoplanet population to EMFs and thus total volatile masses larger than 1\,\%. This occurs when we are looking at planets with mass and radii requiring $>1\%$ volatile mass fraction.  However, in these cases the surface pressure of these planets will still increase as a function of total volatile mass if fugacity effects are accounted for, albeit more slowly than for the ideal gas scenario. Therefore, the effects of non-ideality will not affect our results substantially.
}

\subsubsection{Atmospheric Climate Module}
\label{subsubsec:climatemodule}

Once the composition of the atmosphere has been determined, the pressure-temperature structure of the atmosphere is calculated using the 1D radiative-convective climate model \texttt{AGNI} \citep{Nicholls2025,Nicholls2025b,nicholls_thesis_2026} at each time step. \texttt{AGNI} solves for radiative-convective equilibrium in the atmosphere by optimising its pressure temperature profile such that net radiative, convective, latent heat and sensible heat fluxes lost across each layer of the atmosphere are minimised. It determines this physical solution for the temperature profile using a damped Newton-Raphson optimisation method. This method conserves energy fluxes both globally and locally, while permitting a net heat flux to be carried by the atmosphere. \texttt{AGNI} parametrises convection in 1D using mixing length theory, under the Schwarzschild criterion with an asymptotic mixing length \citep{Vitense1953,Robinson2014,Lee2024}. This parametrisation allows dry convective fluxes to be calculated directly at each layer of the atmosphere as a function of temperature (see equation 2 in \cite{Nicholls2025}). The temperature structure and convective fluxes are relatively insensitive to the mixing length \citep{Joyce2023}.

{\textcolor{black}{Radiative fluxes are calculated using the established \texttt{SOCRATES} radiative transfer suite \citep{edwards_efficient_1996,Edwards1996,manners_socrates_2024,Sergeev2023}. \texttt{SOCRATES} computes radiative fluxes by solving the plane-parallel, two-stream radiative transfer equation. We use 48 correlated-$k$ bands \citep{Lacis1991, goody_atmospheric_1989} fitted to line opacity data from the \texttt{DACE} database \citep{Grimm2021}, for wavenumbers ranging from 0 cm$^{-1}$ to 42000 cm$^{-1}$ \citep{Nicholls2025}. The opacity data taken from the DACE database has a spectral resolution of 0.01\,cm$^{-1}$. Rayleigh scattering of the longwave and shortwave radiation streams and collisionally-induced continuum absorption are also included. We neglect the potential radiative effects of aerosols and clouds for these H$_2$-dominated atmospheres, and assume that the atmosphere is isochemical -- the molecular composition of the atmosphere is vertically constant, set by the lower-boundary condition calculated by solubility-thermochemical equilibrium at each iterative time-step.} This approach is in line with previous work \citep{Lichtenberg2021,Boer2025,Nicholls2025d}. Real gas equations of state are used to calculate the atmosphere height structure: \citet{Haldemann2020} for water; \citet{Saumon1995} for hydrogen; and the Van der Waals equation of state for the other species. The atmosphere consists of discrete pressure layers from 10$^{-5}$ bar down to the total surface pressure as calculated by the \texttt{CALLIOPE} outgassing code. We do not include the effects of atmospheric escape in this work (although see section \ref{subsec:escapejustifiaction}).

\textcolor{black}{The EMF is parametrised, with no physical relationship between between the EMF and the mass of the iron-silicate interior set by atmospheric escape. While the EMF will be related to the interior mass by atmospheric escape \citep{Rogers2021}, this is not an exact relation: it is merely an upper limit \citep{Tang2024}. The circumstances of planet formation, namely the location of planet formation within the disc, disc mass and chemistry, magma solubility and chemistry, will determine the EMF at birth \citep{Bean2021}. The mass of the rocky core can also affect the envelope mass fraction, and thus planetary radius, through solubility, however this effect is minor (see figures \ref{fig:massradiusplot}, \ref{fig:oxygenfugacityatmmasses} and \ref{fig:carbonstudyatmmasses}). Regardless of the relationship between interior mass and EMF, it is still informative to treat these as independent parameters in an evolutionary modelling framework, as these properties affect planetary evolution in different ways. Interior mass affects the thermal state of the mantle via pressure effects in the deep mantle (see section \ref{subsec:planetmass}), while EMF affects the surface temperature and thus the melt state \citep{Lichtenberg2021,Innes2023,Rigby2024,Tang2024}. In any case, our results will not be affected by our neglecting atmospheric escape or the relationship between interior mass and atmospheric escape (see section \ref{subsec:escapejustifiaction}).}

In this study, we adopt blackbody stellar emission spectra defined by an effective temperature and bolometric luminosity, and therefore we do not account for stellar evolution. While libraries of semi-empirical stellar spectra do exist \citep[e.g.,][]{France2016}, they lack the resolution in stellar effective temperature that we require in this study. Libraries of synthetic stellar spectra also exist, but many lack synthetic spectra for M-type stars \citep{Wheeler2023}. In  appendix section \ref{apsec:blackbodytests}, we conduct sensitivity tests comparing modelled atmosphere PT profiles calculated using semi-empirical MUSCLES spectra against those calculated using blackbody spectra. Calculated surface temperatures are only sensitive to the stellar spectrum for the M-star regime due to optical-wavelength differences, where MUSCLES is poorly constrained due to ISM attenuation \citep{Wilson_TheMegaMU_2024,Youngblood_THEMUSCLES_2016}. \textcolor{black}{Also, our results are likely insensitive to the XUV region of the stellar spectrum due to the fact that we do not model atmospheric escape.} We parametrise the stellar radius, and correspondingly the stellar luminosity, as a function of the stellar effective temperature ($T_{\text{eff}}$) using the empirical mass-radius relation from \cite{Demircan1991} ($R_\star\propto M_\star^{0.945}$), the linear mass-luminosity relation from \cite{Eker2015} ($L_\star\propto M_\star^{4.04}$) and the Stefan-Boltzmann law. This also allows us to calculate the stellar instellation flux ($F_{\text{ins}}$) as a function of $T_{\text{eff}}$ and the orbital semi-major axis $a_{\text{orb}}$. We assume an eccentricity of 0 for all simulated planets, given the close-in orbits of most detected sub-Neptunes.

Within our time-stepping loop, the net flux at the topmost layer of the atmosphere calculated by \texttt{AGNI} energy conservation is used to determine the flux upper-boundary condition for \texttt{SPIDER} in the next iteration of the evolutionary model. \texttt{PROTEUS} repeats this iterative process until either the average melt-fraction of the whole mantle decreases below 1~wt\%, corresponding to solidification, or the net flux leaving the planet decreases below 0.2~W~m$^{-2}$, corresponding to global flux balance and a steady state magma ocean. All the quantities presented in this work, excluding those presented in section \ref{subsec:timescalestudy}, correspond to the time at which the simulated planet achieves either mantle solidification or net flux balance.

\subsection{Planetary Classification Criteria}
\label{subsec:gasdwarfwaterworlddefinitions}

\textcolor{black}{To provide clarity around the terms used in this work and the scope of our findings, we outline and provide justification for the criteria we use to define `gas dwarf' exoplanets and `water world' exoplanets. These criteria are contained in table \ref{tab:gasdwarfwaterworldcriteria}.}

\begin{table}
\centering
\begin{tabular}{l|c c}
\hline
 & Gas Dwarf & Water World \\
\hline
Planet Radius ($R_{\oplus}$) & $1.8\!<\!R\!<\!4$ & $1.8\!<\!R\!<\!4$ \\
Total H$_2$O Mass Fraction (\%) & <10 & >10 \\
Total C Mass Fraction (\%) & <10 & <10 \\
Total S Mass Fraction (\%) & <10 & <10 \\
Atmospheric Mass Fraction (\%) & 0.1--10 & >10 \\
Mean Molecular Weight (g mol$^{-1}$) & <3.8 & >3.8 \\
\hline
\end{tabular}
\caption{\textcolor{black}{The criteria used to define `Gas Dwarf' and `Water World' in this work. Any further reference to either of these terms refers to a planet that fulfils all of the respective criteria. We define these criteria not to propose definitions that the community should adopt, but to delineate the parameter space for which our conclusions are valid.}}
\label{tab:gasdwarfwaterworldcriteria}
\end{table}

\textcolor{black}{Given that gas dwarfs and water worlds are both sub-categories within the sub-Neptune classification, we require that both planet types have radii between 1.8\,R$_{\oplus}$ and 4\,R$_{\oplus}$. We define the transition between gas dwarf and water world in terms of total water mass fraction as 10\,\%. This is approximately the water mass fraction at which a super-Earth stripped of a hydrogen-dominated atmosphere would have radii consistent with typical sub-Neptune radii \citep{Rogers2024}. To distinguish gas dwarfs and water worlds from sub-Neptunes that achieve radii between 1.8\,R$_{\oplus}$ and 4\,R$_{\oplus}$ as a result of large carbon or sulphur mass fractions \citep{Bergin2023ApJL,Li2025,Nicholls2025c}, we require that both planet types have total carbon and sulphur mass fractions less than 10\,\%. Based on the findings of \citet{Rogers2021}, who suggest that sub-Neptunes could form with H$_2$-rich atmospheres with EMFs ranging from 0.1--10\,\%, we require that gas dwarfs have EMFs ranging from 0.1--10\,\%. We also require that gas dwarfs have a mean molecular weight less than 3.8\,g mol$^{-1}$. As an example of an atmospheric composition this translates to, this is the mean molecular weight of an atmosphere composed of 85\,\% H$_2$, 10\,\% CH$_4$ and 5\,\% H$_2$O. This definition is similar to \cite{Benneke2024}, who define gas dwarfs as planets with an Earth-like interior and a low-mean molecular weight atmosphere (<3\,g mol$^{-1}$). Similar definitions for a gas dwarf are widely used in the literature surrounding sub-Neptunes \citep{Buchhave2014,Zeng2019,Chouqar2020,Gao2023,Rigby2024,Rogers2025,Bower2025}.}

\textcolor{black}{Any reference to a `gas dwarf' or a `water world' in this work refers to a planet the meets all of the criteria we use for either definition. We define these criteria not to propose definitions for gas dwarf and water world that the community should adopt, but to delineate the parameter space for which our conclusions are valid.}




\section{The Solidification Shoreline}
\label{sec:solidshoreline}

In order to determine the fraction of the gas dwarf population with magma oceans, we seek to divide the parameter space relevant for gas dwarfs into two regions: one in which our model predicts that gas dwarfs will reach a thermal steady state that permits a surface magma ocean, and one in which they are predicted to cool to the point of complete mantle solidification. This concept is analogous to the `Habitable Zone' \citep{Hunag1960,Kasting1993,Kopparapu2013,Turbet2023}, which seeks to divide the host star effective temperature-instellation flux parameter space into a region where liquid water is stable on the surface of a planet and a region where liquid water is unstable due to freezing. Another example of a division of a multi-dimensional parameter space into two distinct regions is the `Cosmic Shoreline' \citep{Zahnle2017,Lustig2019,Moran2023,Chatterjee2024}. The cosmic shoreline is a tool to separate the escape velocity-instellation flux parameter space into a region where planets are expected to have atmospheres and a region where planets are not expected to have atmospheres. In a similar vein to studies that have developed the habitable zone and cosmic shoreline as a tool, we seek to determine the `solidification shoreline' that separates gas dwarfs with permanent magma oceans and those with solidified mantles.

To determine the location of the solidification shoreline separating gas dwarfs that will solidify and those that will remain permanently molten, we must first choose a parameter space on which the solidification shoreline is defined. Given that the melt fraction of the mantle is strongly dependent on the surface temperature \citep{Jeroen2015,MONTEUX2016}, and that the surface temperature correlates with the planet's equilibrium temperature \citep{Kasting1993,DelGenio2019,Herath2024}, it is natural to choose the instellation flux as one of the parameters. Estimates of the habitable zone are strongly dependent on the effective temperature of the host star, given that the bond albedo of the planet depends on the wavelength of the incident starlight \citep{Kasting1993,Kopparapu2013}. It is unclear how sensitive the thermal steady states of gas dwarfs are to the stellar effective temperature when the instellation flux is held constant. 

\cite{Tang2024} show that the envelope mass fraction (EMF) of a gas dwarf has a significant effect on the cooling of the mantle. However, the EMF is more difficult to infer from observations of sub-Neptunes than the host star effective temperature and the instellation flux. It is much easier to place a planet in instellation flux-stellar effective temperature parameter space without requiring more detailed observation of the system. The solidification shoreline heuristic therefore allows us to identify planets that merit further investigation for magma ocean phenomena with only minimal system information. 

The composition of a planet's envelope will also influence the opacity of the atmosphere, and thus the cooling of the mantle \citep{Sossi2020,Gaillard2022,Nicholls2024,Tang2024}. However, given that our gas dwarf definition requires an atmosphere with $\mu<$3.8\,g mol$^{-1}$, the atmospheric composition can only vary within a limited parameter space (see section \ref{subsec:mmwconstraints} for a more detailed discussion of this issue). Therefore, within our gas dwarf paradigm, envelope mass fraction will have a greater influence on the location of the solidification shoreline than envelope composition. For this reason and those discussed above, we choose to define the solidification shoreline in $T_{\text{eff}}$ vs. $F_{\text{ins}}$ parameter space, while also studying the effect of varying the EMF on the location of the shoreline.


\subsection{Determining the Solidification Shoreline as a Function of Envelope Mass Fraction}
\label{subsec:solidshrelineemf}

We perform simulations with \texttt{PROTEUS} across the $T_{\text{eff}}$-$F_{\text{ins}}$-EMF parameter space to estimate the location of the solidification shoreline. We sample stellar effective temperatures from 2900~K to 6300~K, covering M, K, G and F stars, similar to \cite{Turbet2023}. Given that \cite{Tang2024} investigate the cooling of sub-Neptunes for instellation fluxes spanning several orders of magnitude, and that we are interested in instellation fluxes low enough to result in solidification, we sample instellation fluxes from 10$^{-1}$$F_{\oplus}$ to 10$^{3}$$F_{\oplus}$. We adopt a value of 1361~W~m$^{-2}$ for $F_{\oplus}$, using the IAU definition of the solar luminosity and the astronomical unit. Given that gas dwarfs are expected to form with EMFs ranging over several orders of magnitude \citep{Wolfgang2015,Ginzburg2016,Burn2024,Tang2024}, we perform simulations with EMFs of 0.1\%, 0.3\% and 1\%. \textcolor{black}{While gas dwarfs could form with an EMF as high as 10\% \citep{Rogers2021}, the trend in figure \ref{fig:solidificationshoreline} shows that any shorelines for EMFs greater than 1\,\% would be located at instellation fluxes lower than the smallest instellation fluxes of the detected sub-Neptune population. Therefore, any currently detected gas dwarfs with EMFs greater than 1\,\% would have permanent magma oceans.}

Atmospheric escape will mean that the EMF inferred for a given planet today will underestimate the EMF that the planet will have had in its past. We do not model atmospheric escape, and instead compare the model to planets by fixing the EMF at a value consistent with that inferred at their present.  The result of this is that the simulated planet will cool more quickly than if its envelope were initialised more massive and allowed to lose mass down to its present state. Inclusion of atmospheric escape would therefore globally shift the solidification shoreline to lower instellation fluxes, for a given final EMF, making our results a conservative estimate of whether gas dwarfs will be molten.

In our baseline simulations, we set the C/H and N/H mass ratios of the bulk volatile inventory in the mantle to 0.32 and 0.09 respectively. This is equivalent to the C/H and N/H mass ratios corresponding to 100$\times$ that of the sun, i.e., `100$\times$ solar metallicity' \citep{Asplund2021}. We use an oxygen fugacity of $\Delta \text{IW}=-3$, to ensure that the resulting atmosphere is H$_2$-dominated and thus consistent with the gas dwarf classification, and a planet mass of 5$M_{\oplus}$. The volatile hydrogen mass fraction, relative to the mass of the mantle, is set to the value needed to achieve the required EMF, assuming that the entire volatile budget is in the atmosphere rather than the melt. \textcolor{black}{The planetary radius of our fiducial model, i.e., the model with the fiducial parameter values outlined in table \ref{tab:constantparameters}, is 2.16\,R$_{\oplus}$.}

Using the grid of \texttt{PROTEUS} simulations spanning the $T_{\text{eff}}$-$F_{\text{ins}}$ parameter space, we determine the instellation flux at which the thermal steady state of the planet transitions from a permanent magma ocean to a solidified mantle; i.e., the instellation flux corresponding to the solidification shoreline, $F_{\text{shore}}$, for each stellar effective temperature, for each EMF. We then fit $T_{\text{eff}}$ as a function of log$_{10}(F_{\text{shore}}$) to a second order polynomial using least-squares regression for each EMF,

\begin{equation}
    F_{\text{shore}}=a_0T_{\text{eff}}^2+a_1T_{\text{eff}}+a_2.
    \label{equ:shorelineanalyticalfit}
\end{equation} The coefficients for the polynomial fits are shown in table \ref{tab:polycoeffs}. Any further reference to the location of the solidification shoreline refers specifically to these analytical functions. We note that equation \ref{equ:shorelineanalyticalfit} is an approximation: the threshold flux for magma ocean solidification will in reality depend on more parameters than simply the stellar effective temperature and the envelope mass fraction (see section \ref{sec:secondaryparameters}). However, given the restrictions on atmospheric composition imposed by the gas dwarf paradigm and our results in section \ref{sec:secondaryparameters}, stellar effective temperature and envelope mass fraction are the most significant parameters that determine the threshold instellation flux for planetary solidification.

\begin{table}
\centering
\begin{tabular}{lccc}
\hline
\textbf{Envelope Mass Fraction (\%)} & \textbf{a$_0$} (K$^{1/2}$) & \textbf{a$_1$} (K) & \textbf{a$_2$} (K) \\
\hline
0.1  & 2.548$\times$10$^{-8}$ & -3.587$\times$10$^{-4}$ & 2.557 \\
0.3  & 5.042$\times$10$^{-8}$ & -5.607$\times$10$^{-4}$ & 2.605 \\
1.0  & 3.909$\times$10$^{-8}$ & -3.951$\times$10$^{-4}$ & 1.006 \\
\hline
\end{tabular}
\caption{Polynomial coefficients for the analytical fit to the instellation flux at the solidification shoreline as a function of stellar effective temperature ($F_{\text{shore}}=a_0T_{\text{eff}}^2+a_1T_{\text{eff}}+a_2$). Coefficients are shown for different envelope mass fractions.}
\label{tab:polycoeffs}
\end{table}

\begin{figure*} 
    \centering
    \begin{subfigure}[b]{0.49\textwidth}
        \centering
        \includegraphics[width=\linewidth]{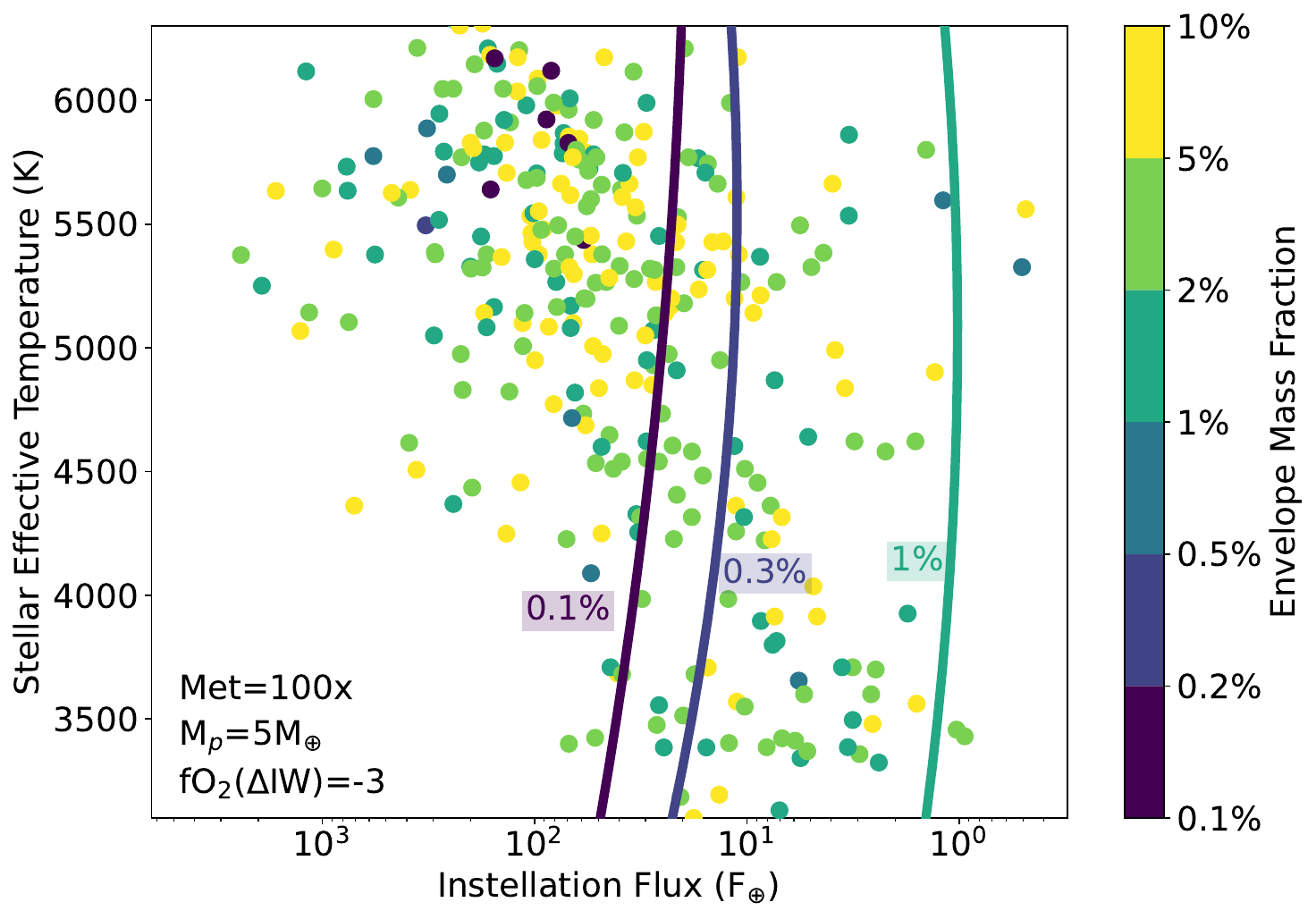}
        \label{fig:first}
    \end{subfigure}
    \begin{subfigure}[b]{0.49\textwidth}
        \centering
        \includegraphics[width=\linewidth]{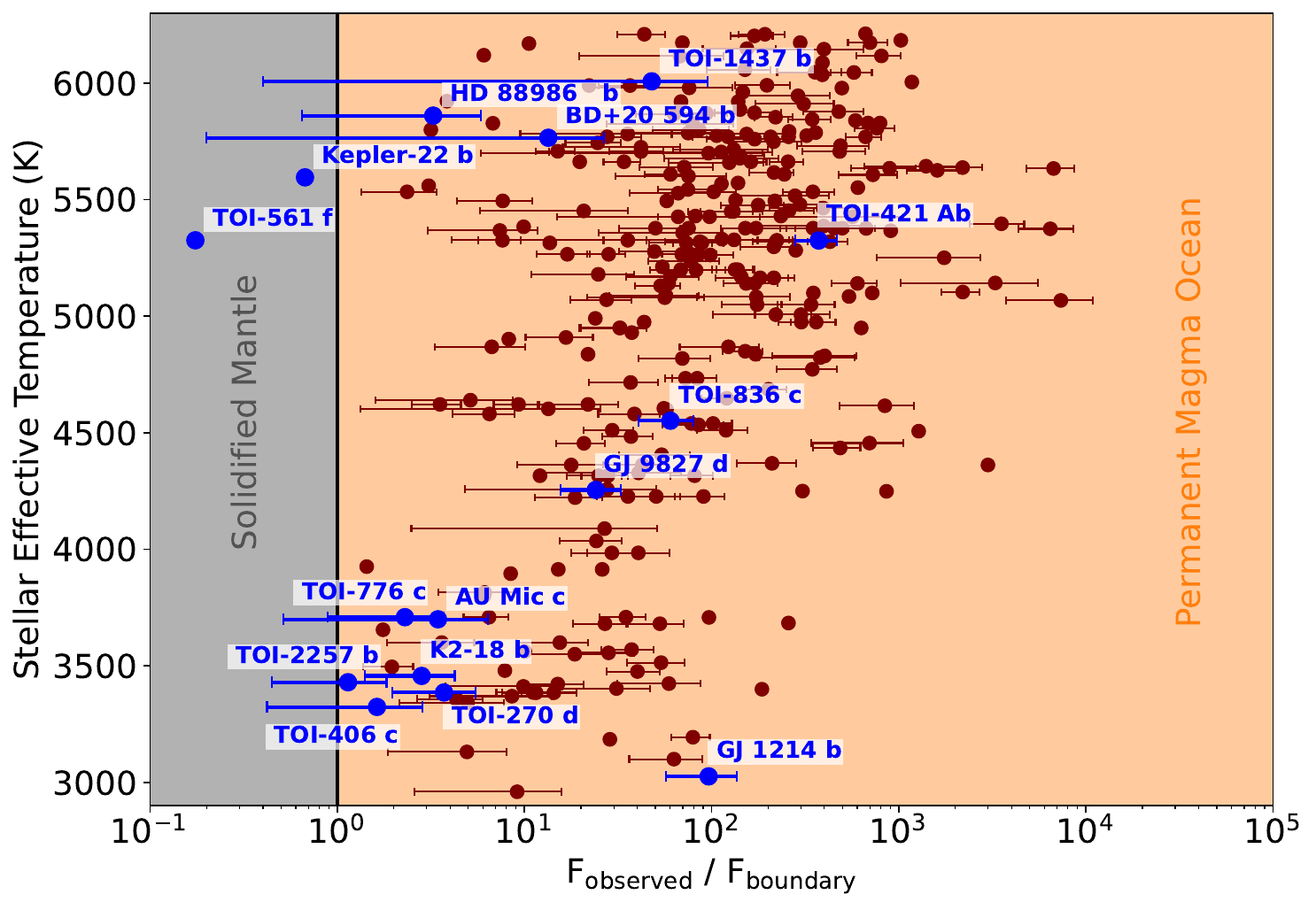}
        \label{fig:second}
    \end{subfigure}
    \caption{\textbf{Left}: Instellation flux at which the thermal steady state of a gas dwarf transitions from a permanent magma ocean to a solidified mantle: the `solidification shoreline'. Planets in the region of the parameter space to the left of the shoreline will have permanent magma oceans, and any planets to the right of the shoreline will have solidified mantles. The solidification shoreline is shown for envelope mass fractions of 0.1\%, 0.3\% and 1\%. The positions of all detected sub-Neptunes in this parameter space are shown as scatter points; their colour corresponds to the EMF calculated using equation 2 from \protect\cite{Lopez2014}. \textbf{Right}: Ratio of the observed instellation flux of detected sub-Neptunes ($F_{\text{observed}}$) to the instellation flux required for them to be on their appropriate solidification shoreline ($F_{\text{boundary}}$) as a function of host star effective temperature. Errors in $F_{\text{observed}}$/$F_{\text{boundary}}$ are propagated from observed masses, host star effective temperature and planet masses. A ratio of F$_{\text{observed}}$/F$_{\text{boundary}}$=1, corresponding to a planet located on the solidification shoreline, is shown. Sub-Neptunes for which there exist published JWST transmission spectra as well as those in the solidified mantle region are shown in blue, and those without JWST transmission spectra are shown in maroon.}
    \label{fig:solidificationshoreline}
\end{figure*}

For a larger EMF, we find that the solidification shoreline is located at lower instellation fluxes (Figure \ref{fig:solidificationshoreline}), implying that sub-Neptunes with a larger EMF are more likely to have a permanent magma ocean for the same instellation flux. This is expected, given that the H$_2$ continuum opacity scales with the square of pressure, implying that a higher surface pressure results in a higher atmospheric opacity, increasing the surface temperature. These EMFs are not large enough to result in surface pressures above the threshold required to induce solidification \citep{Breza2025}.

The solidification shoreline exists at lower instellation fluxes for planets around host stars with a higher $T_{\text{eff}}$ (the slight increase in $F_{\text{shore}}$ with increasing $T_{\text{eff}}$ for $T_{\text{eff}}$>5000\,K and EMF=1\,\% is a by-product of the polynomial fit to the simulation results), though this relationship is weak. This insensitivity of the location of the shoreline with respect to $T_{\text{eff}}$ contrasts with the strong dependence of the location of the habitable zone on stellar effective temperature \citep{Hunag1960,Kasting1993,Kopparapu2013,Turbet2023}. \textcolor{black}{This is a radiative transfer effect: unlike studies that estimate the location of the habitable zone, we do not include opacity contributions from clouds, which scatter or absorb incident starlight depending on its wavelength.} A caveat to this finding is that our stellar spectra sensitivity tests (section \ref{apsec:blackbodytests}) imply that the shoreline may be located at lower instellation fluxes than those suggested by figure \ref{fig:solidificationshoreline} for M-type stars, as a result of our stellar blackbody approximation.

The positions of all detected sub-Neptunes (exoplanets with 1.8$R_{\oplus}<R<$4$R_{\oplus}$) in the $T_{\text{eff}}$-$F_{\text{ins}}$ parameter space are also shown in figure \ref{fig:solidificationshoreline}. Data for the planets are taken from the \url{exoplanet.eu} database \citep{Schneider2011}. The EMFs of the detected sub-Neptunes are estimated using the analytical relationship (equation 2) between EMF, planet radius, planet mass, instellation flux and planet age in \cite{Lopez2014}. While \cite{Lopez2014} state that this analytical law should not be used to predict the EMF for specific planets, we believe that it is a sufficient approximation for our population study. Also, while the equation in \cite{Lopez2014} is only valid for pure H/He atmospheres, if it were adapted to account for the carbon- and nitrogen-enriched atmospheres in our simulations, it would only lead to higher predicted EMFs due to larger atmospheric mean-molecular weights. This increase in EMF would only make the conclusion of most gas dwarf sub-Neptunes being molten more robust. We calculate the core radius using equation 2 in their paper, and we neglect the radius of the radiative atmosphere, given that the instellation fluxes corresponding to the solidification shorelines are much lower than the largest instellation fluxes considered in their work. We assume that all planets have an age of 5 Gyr and assume an uncertainty in the age of 3 Gyr.

To quantify the degree to which observed sub-Neptunes are either close to or far away from their appropriate solidification shoreline, we calculate the instellation flux at which the expected thermal steady state transitions from solidified mantle to a permanent magma ocean for each observed sub-Neptune, $F_{\text{boundary}}$. We calculate $F_{\text{boundary}}$ for an observed sub-Neptune by predicting the instellation flux it would have at the solidification shoreline corresponding to its estimated EMF based upon the solidification shorelines shown in figure \ref{fig:solidificationshoreline}, using linear interpolation and extrapolation. The right panel of figure \ref{fig:solidificationshoreline} shows the ratio of the observed instellation flux and the instellation flux at the shoreline, $F_{\text{observed}}$/$F_{\text{boundary}}$, for each observed sub-Neptune. A value greater than 1 for this ratio implies the planet will have a permanent magma ocean in thermal steady state. We determine the uncertainty for this ratio by propagating the errors for the observed stellar effective temperatures, semi-major axes and the planet masses. We see that, for the majority of the observed sub-Neptunes (98\%), $F_{\text{observed}}$/$F_{\text{boundary}}\gg1$, indicating that they are very far from their respective solidification shoreline in the $T_{\text{eff}}$-$F_{\text{ins}}$ parameter space (figure \ref{fig:solidificationshoreline}, right panel). Only a handful of sub-Neptunes could potentially exist within the `solidified mantle' region of the parameter space, given the uncertainty in their respective flux ratios. Therefore, the majority of detected sub-Neptunes will have permanent magma oceans if they are (1) consistent with our gas dwarf criteria, and (2) possess a present day EMF equal to or greater than that which they had immediately after their boil-off phase.

While the EMF, host star effective temperature and instellation flux are the most significant and/or easily observable planetary parameters that determine whether a gas dwarf will solidify, there are other parameters that could influence its thermal steady state. If these parameters were to vary from our fiducial model for a given gas dwarf, this would result in a hotter or colder interior. The given gas dwarf would then require either a greater or lower instellation flux to be located at the boundary between solidification and a permanent magma ocean. Therefore, the location of the solidification shoreline may depend on these secondary parameters; namely the oxygen fugacity of the magma ocean, the C/H ratio of the bulk volatile inventory, and the planet mass. This motivates a study of the effect of varying these secondary parameters on the location of the solidification shoreline, similar to work done by \cite{Ji2025}, who demonstrated the sensitivity of the cosmic shoreline to parameters other than escape velocity and instellation flux.

\section{Effects of Secondary Parameters on the Solidification Shoreline}
\label{sec:secondaryparameters}

To understand the effect of the oxygen fugacity of the magma ocean, the C/H ratio of the bulk volatile inventory and the planet mass on the location of the solidification shoreline in the $T_{\text{eff}}$-$F_{\text{ins}}$ parameter space, we perform simulations with \texttt{PROTEUS} in which these parameters are independently varied. In the first and second of these secondary parameter studies, we set the EMF to 0.1\%, for computational feasibility. Also, for the first and second studies, we use a stellar effective temperature of 4500~K and an instellation flux of 27\,$F_{\oplus}$, to ensure that these simulations are located near the solidification shoreline in the $T_{\text{eff}}$-$F_{\text{ins}}$ parameter space. 

To quantify the effect of varying the parameter in question on the location of the solidification shoreline, we record the global melt fraction of the mantle as well as the net atmospheric flux when the planets achieve their thermal steady state: either a solidified mantle or a permanent magma ocean. If varying the parameter results in an increased melt fraction at the end of the thermal evolution for planets with permanent magma oceans, then these planets will require lower instellation fluxes to achieve solidification. Conversely, if varying the parameter increases the net atmospheric flux at the end of the thermal evolution for planets that solidify, then these planets require higher instellation fluxes to achieve flux balance and thus a permanent magma ocean.

\subsection{Oxygen Fugacity of the Magma Ocean}
\label{subsec:oxygenfugacity}

The redox state of the mantle exerts significant influence over the speciation of an atmosphere overlaying a magma ocean \citep{HIRSCHMANN2012,Sossi2020,Gaillard2022}. Highly reducing mantles result in H$_2$ dominated atmospheres, with CH$_4$ and CO as the main secondary constituents, whereas more oxidising mantles result in H$_2$O-CO$_2$ dominated atmospheres. Given that the radiation absorption cross-sections of these molecules differ over the wavelength regime in which the host star is most luminous (figure \ref{fig:molecularopacities}), different atmospheric compositions affect the greenhouse heating of the surface induced by the atmosphere, which will affect the thermal evolution of the planet's magma ocean \citep{Nicholls2024}. While the atmospheric composition, and thus the oxygen fugacity, can only vary to a limited degree within our gas dwarf constraints (see section \ref{subsec:mmwconstraints}, it is still necessary to understand how the location of the solidification shoreline will change for gas dwarfs with more- or less-oxidising mantles. To investigate the effect of mantle oxidation state on the thermal steady state of gas dwarfs, we vary the oxygen fugacity relative to the iron-w\"ustite buffer between $\Delta{\rm IW}=-5$ and $\Delta{\rm IW}=+5$. This range was chosen to cover the transition between an H$_2$-dominated and an H$_2$O-dominated atmosphere.

\begin{figure}
    \includegraphics[width=0.49\textwidth]{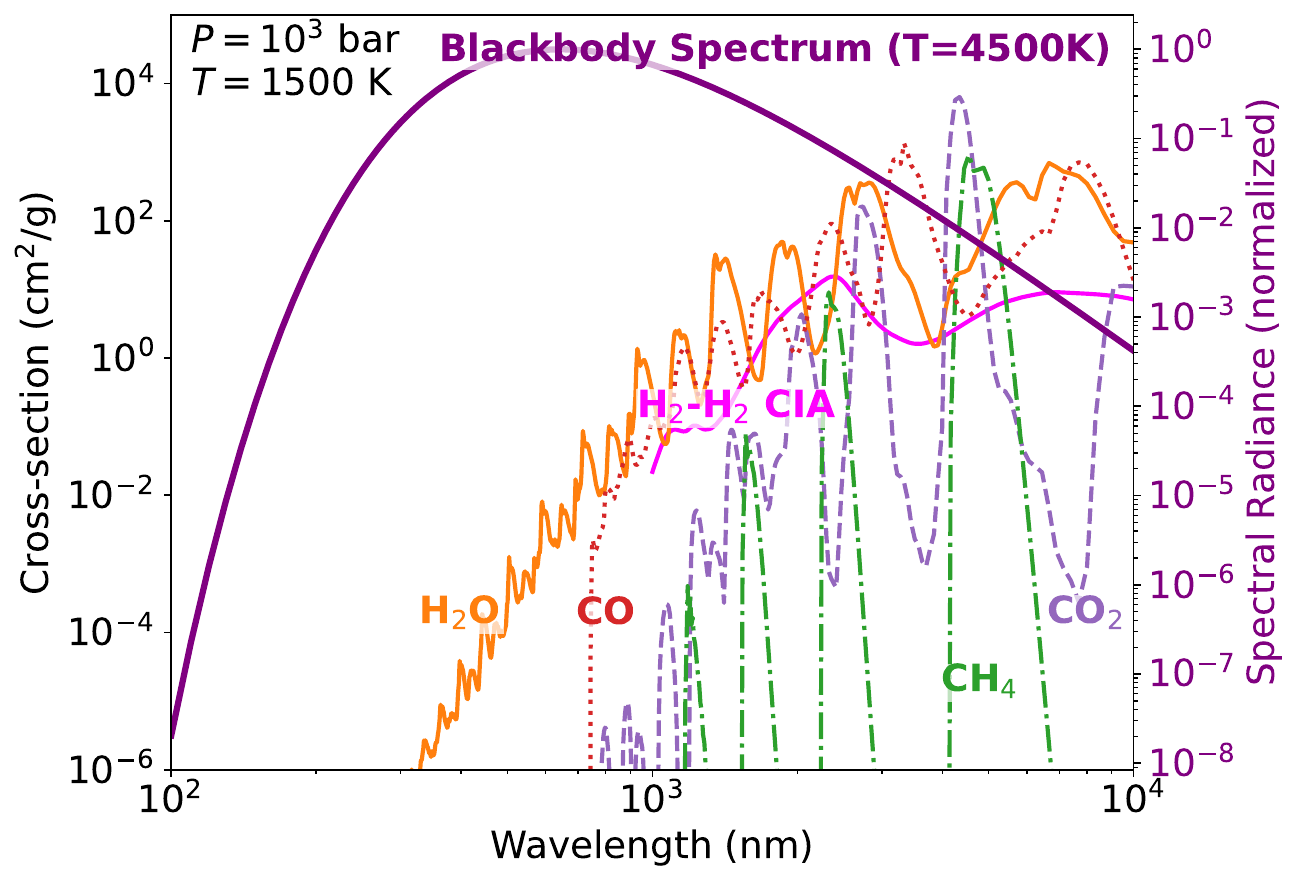}
    \caption{Absorption cross-sections of key species as a function of wavelength for $P=10^{3}$\,bar and $T=1500$\,K (i.e., the pressure and temperature expected near the base of the atmosphere of a gas dwarf). The cross-section due to H$_2$-H$_2$ collisionally induced absorption is also shown. The blackbody spectrum corresponding to an effective temperature of $T=4500$\,K (i.e., an effective temperature typical of a K-type star) is shown. Absorption cross-sections are taken from the \texttt{DACE} database \citep{Grimm2015,Grimm2021} and cross-sections for H$_2$-H$_2$ collisionally induced absorption are taken from the HITRAN database \citep{Gordon2022}.}
    \label{fig:molecularopacities}
\end{figure}

More oxidising mantles lead to a larger global melt fraction and lower net atmospheric flux (figure \ref{fig:oxygenfugacity}) at the end of the thermal evolution of the planet. This result is due to a larger H$_2$O mixing ratio in the atmosphere (figure \ref{fig:oxygenfugacity}); the opacity of H$_2$O over the relevant wavelength regime is much greater than that due to H$_2$-H$_2$ collision-induced absorption (figure \ref{fig:molecularopacities}). The higher opacity results in an increased downward longwave radiative flux at the bottom of the atmosphere (figure \ref{fig:oxygenfugacity}), implying greater heating of the surface by the atmosphere and an increased melt fraction. However, the increase in melt fraction reaches a maximum at a value of $\Delta{\rm IW}=0$, after which point the melt fraction decreases with increasing oxygen fugacity. This change in behaviour is due to the replacement of CO with CO$_2$ as the secondary constituent of the atmosphere, which has a smaller absorption cross-section over the relevant wavelength regime than CO (figure \ref{fig:molecularopacities}). The smaller absorption cross-section diminishes the greenhouse effect, leading to a lower global melt fraction. 

These results imply that gas dwarfs with more oxidising mantles are more likely to end their thermal evolution with a permanent magma ocean than a solidified mantle, so the solidification shoreline will be pushed to lower instellation fluxes for these planets. This effect only applies, however, to planets with volatile C/H and N/H ratios corresponding to 100$\times$ that of the solar abundances. The oxidation state of the mantle could have a different effect on the thermal steady state of gas dwarf under different volatile carbon endowments. In the next section, we investigate the effect of the C/H ratio of the bulk volatile inventory on the location of the solidification shoreline.

\begin{figure*} 
    \centering
    \begin{subfigure}[b]{0.49\textwidth}
        \centering
        \includegraphics[width=\linewidth]{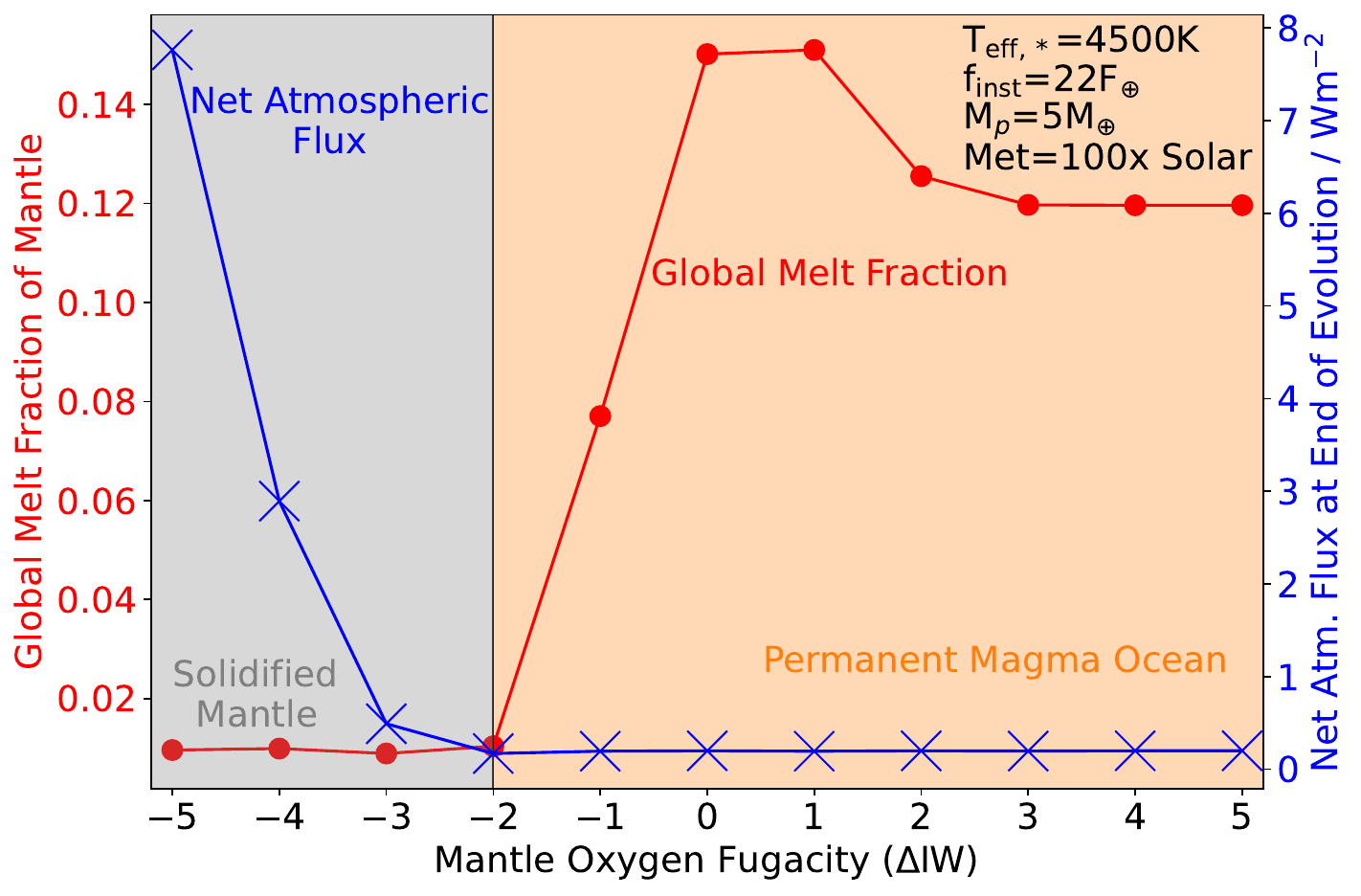}
        \label{fig:first1}
    \end{subfigure}
    \begin{subfigure}[b]{0.49\textwidth}
        \centering
        \includegraphics[width=\linewidth]{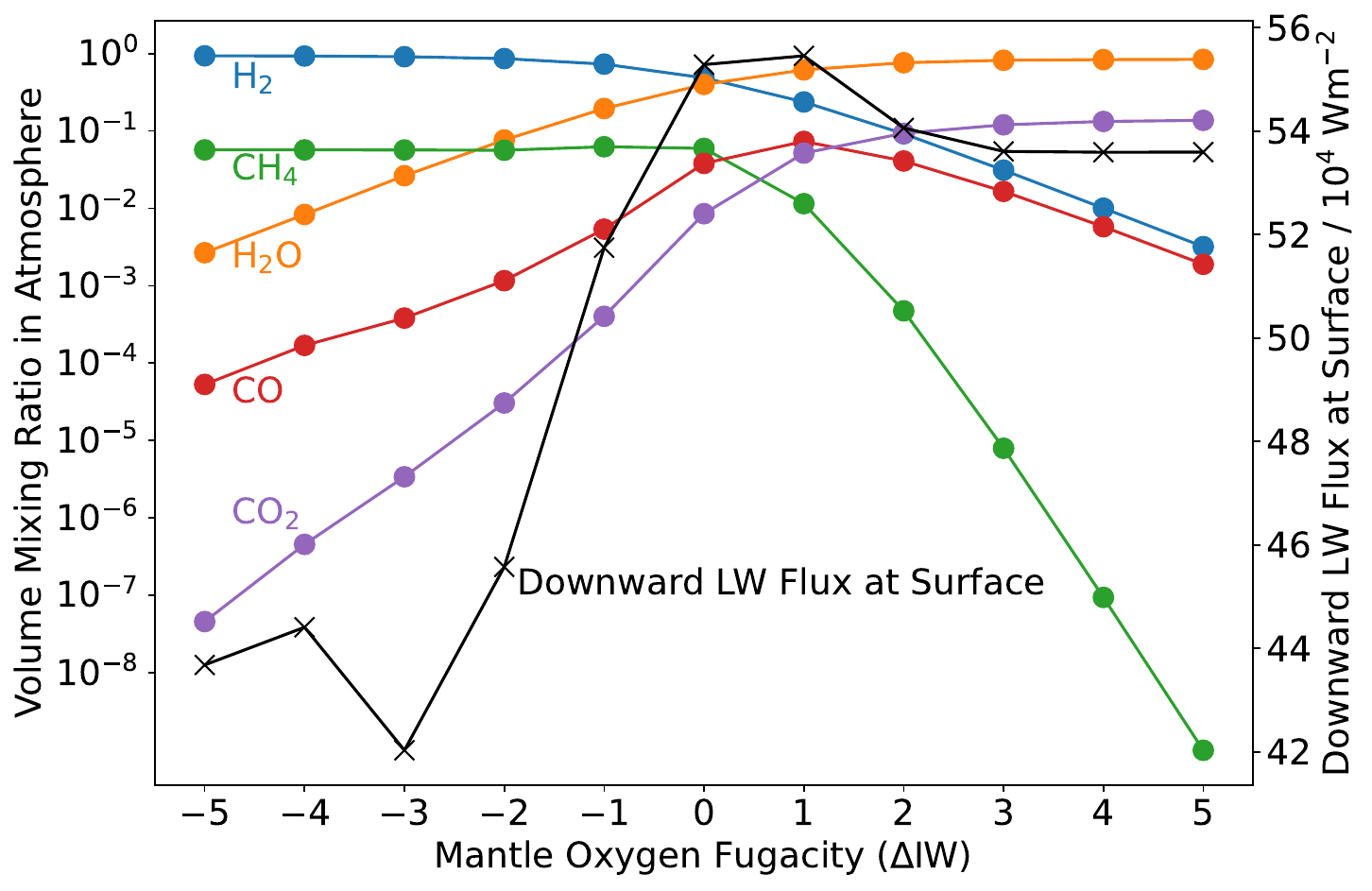}
        \label{fig:second2}
    \end{subfigure}
    \caption{\textbf{Left}: Global melt fraction of the mantle (red dots) as well as the net atmospheric flux at the end of the planet's thermal evolution (blue crosses) as a function of oxygen fugacity of the magma ocean. All simulations with $\Delta{\rm IW} < -2$ achieve mantle solidification, and all simulations with $\Delta{\rm IW} > -2$ end their evolution with a permanent magma ocean. The stellar effective temperature, instellation flux, planet mass and metallicity used in these simulations are also shown. The instellation flux was chosen such that the planet was located at the transition from a permanent magma ocean to a solidified mantle in the $T_{\text{eff,*}}$-$F_{\text{ins}}$ parameter space. \textbf{Right}: Volume mixing ratios of key species in the atmosphere as a function of mantle oxygen fugacity ($\Delta$IW). The downward longwave radiative flux at the base of the atmosphere, which quantifies the rate of surface heating due to the greenhouse effect of the atmosphere, as a function of mantle oxygen fugacity is also shown.}
    \label{fig:oxygenfugacity}
\end{figure*}

\subsection{C/H Ratio of the Bulk Volatile Inventory}
\label{subsec:carbonbudget}

While we have shown that the redox state of the mantle strongly influences the greenhouse heating of the planetary surface for an H$_2$-dominated atmosphere with 100$\times$ solar metallicity, gas dwarfs may have atmospheres that are either carbon-rich or carbon-poor compared to this value. The planet's overall carbon endowment at formation may vary due to the metallicity of the host star \citep{Fortney2012} or the location of the planet in the protoplanetary disc \citep{Li2021}. There is also the possibility of sub-Neptunes with interiors rich in organic molecules; i.e., `soot worlds' \citep{Bergin2023ApJL,Li2025}. Large carbon budgets would result in more volatile carbon outgassing from the magma into the atmosphere. Given the role of molecules such as CH$_4$, CO and CO$_2$ in greenhouse heating of planetary surfaces \citep{Wordsworth2013}, the thermal steady state of gas dwarfs and thus location of the solidification shoreline could depend on the intrinsic carbon budget of the planet, even within the constraints on atmospheric composition imposed by our gas dwarf constraints (see section \ref{subsec:mmwconstraints}). We vary the C/H ratio of the bulk volatile inventory in our model to investigate this dependence.

The volatile C/H ratio of the bulk volatile inventory is varied from 0.1 to 10, encompassing C/H ratios from 100$\times$ solar metallicity up to the bulk-earth C/H ratio \citep{Wang2018}. We vary the C/H ratio at two fixed values of the oxygen fugacity: $\Delta{\rm IW}=-4$ and $\Delta{\rm IW}=+4$, which allows us to explore the speciation of the atmosphere as the C/H ratio is varied for both an oxidising and reducing mantle. We omit nitrogen chemistry in these simulations in order to isolate the effects of carbon chemistry. While this method conserves the total mass of volatile carbon plus hydrogen in the system, it does not conserve the total mass of the atmosphere, given that we do not use a fixed volatile oxygen budget in our simulations. However, the atmospheric masses in these simulations are all consistent to within a factor of 4 (see figures A\ref{fig:oxygenfugacityatmmasses} and A\ref{fig:carbonstudyatmmasses}). Therefore, we do not expect these inconsistent atmospheric masses to affect our findings.

\begin{figure*} 
    \centering
    \begin{subfigure}[b]{0.49\textwidth}
        \centering
        \includegraphics[width=\linewidth]{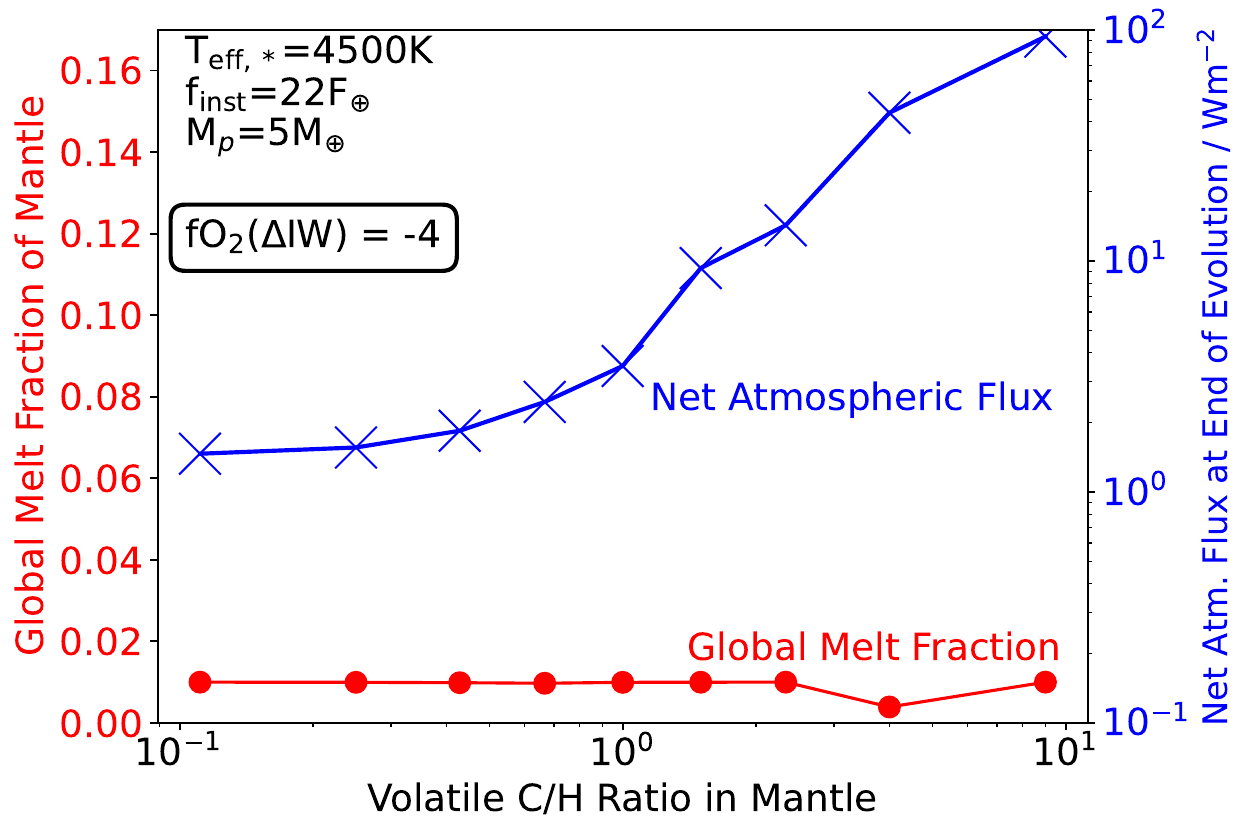}
        \label{fig:first3}
    \end{subfigure}
    \begin{subfigure}[b]{0.49\textwidth}
        \centering
        \includegraphics[width=\linewidth]{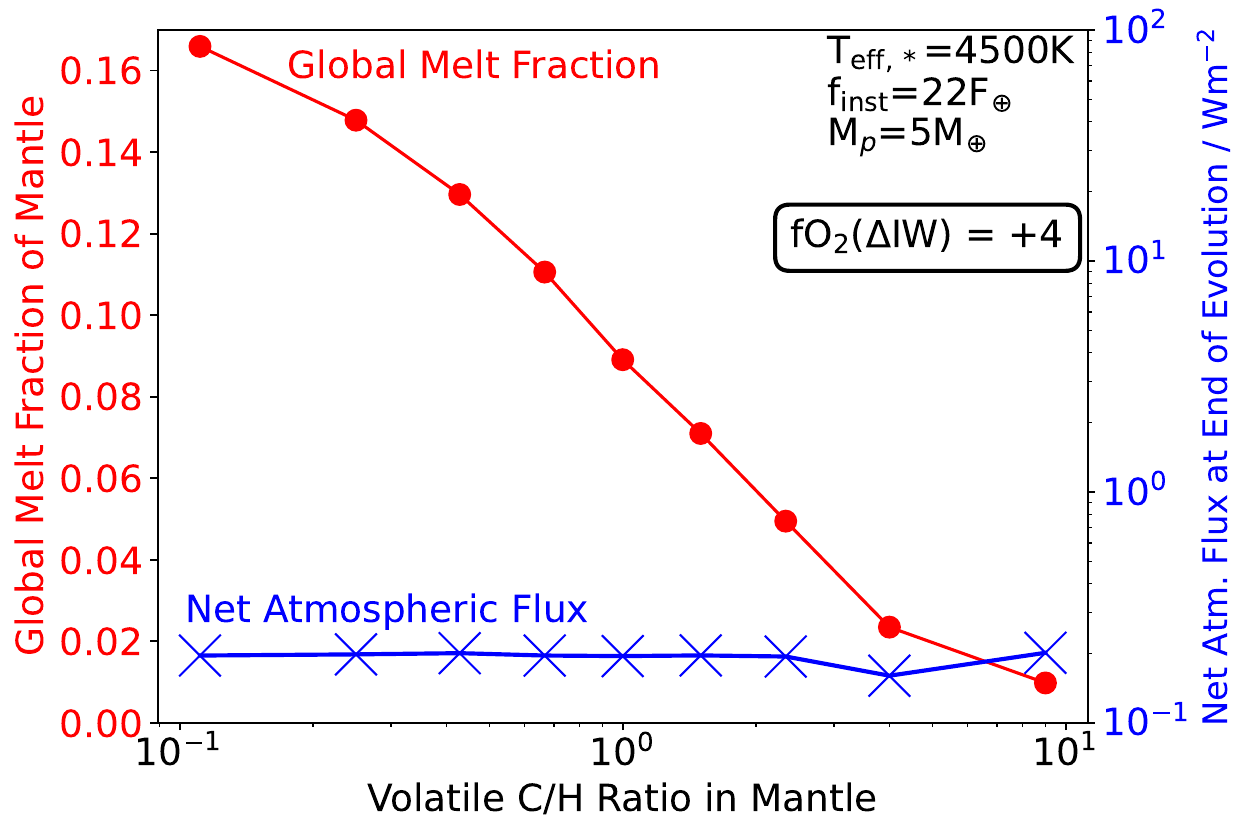}
        \label{fig:second3}
    \end{subfigure}
    \caption{Global melt fraction of the mantle (red dots) as well as the net atmospheric flux at the end of the planet's thermal evolution (blue crosses) as a function of the C/H ratio of the bulk volatile inventory. Results are shown for simulations with a mantle oxygen fugacity of $\Delta{\rm IW}=-4$ (left panel) as well as $\Delta{\rm IW}=4$ (right panel). All of the simulations corresponding to a mantle oxygen fugacity of $\Delta{\rm IW}=-4$ end their evolution with a solidified mantle \textcolor{black}{(hence their constant melt fraction)}, and all of the simulations corresponding to a mantle oxygen fugacity of $\Delta{\rm IW}=4$ end their evolution with a permanent magma ocean \textcolor{black}{(hence their constant net atmospheric flux)}. The stellar effective temperature, instellation flux and planet mass used in these simulations are also listed in the upper corners. The instellation flux was chosen such that the planet was located at the transition from a permanent magma ocean to a solidified mantle in the $T_{\text{eff}}$-$F_{\text{ins}}$ parameter space.}
    \label{fig:carbonstudyresults}
\end{figure*}

\begin{figure*} 
    \centering
    \begin{subfigure}[b]{0.49\textwidth}
        \centering
        \includegraphics[width=\linewidth]{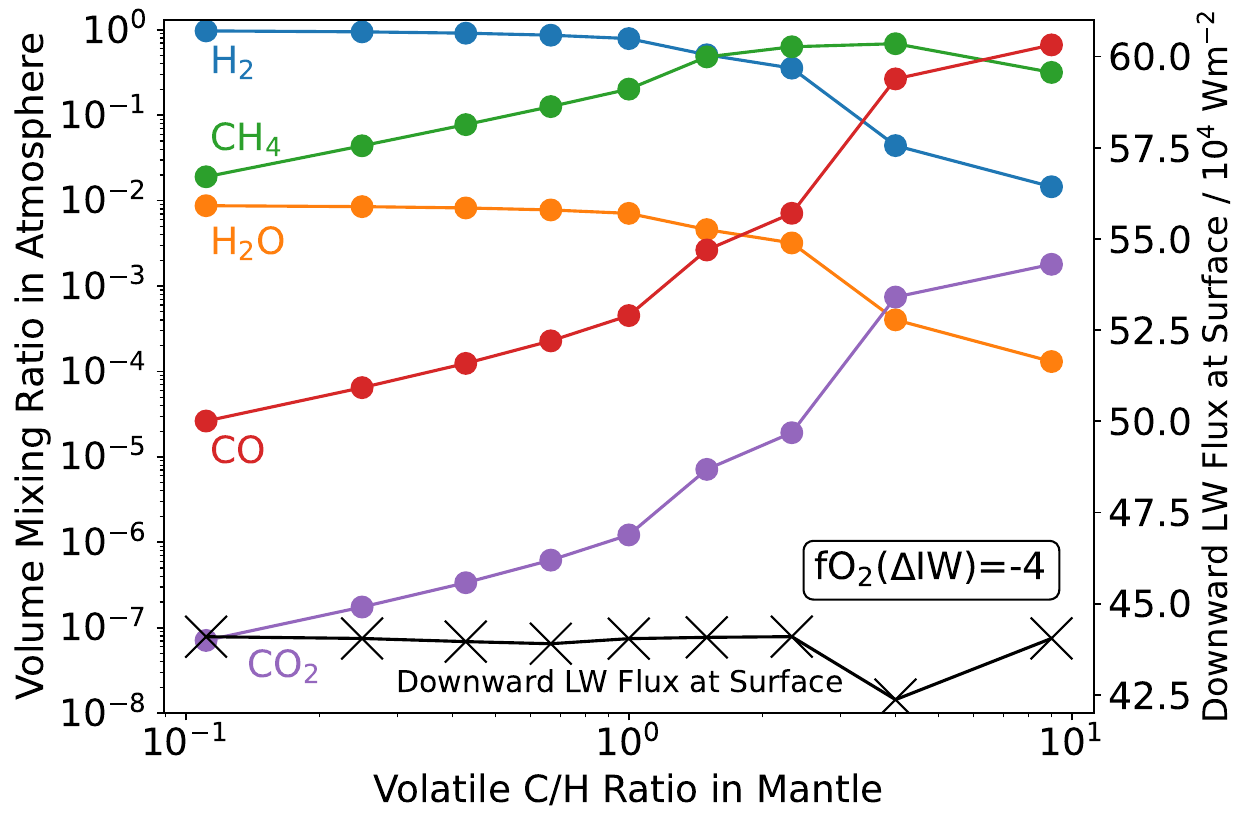}
        \label{fig:first4}
    \end{subfigure}
    \begin{subfigure}[b]{0.49\textwidth}
        \centering
        \includegraphics[width=\linewidth]{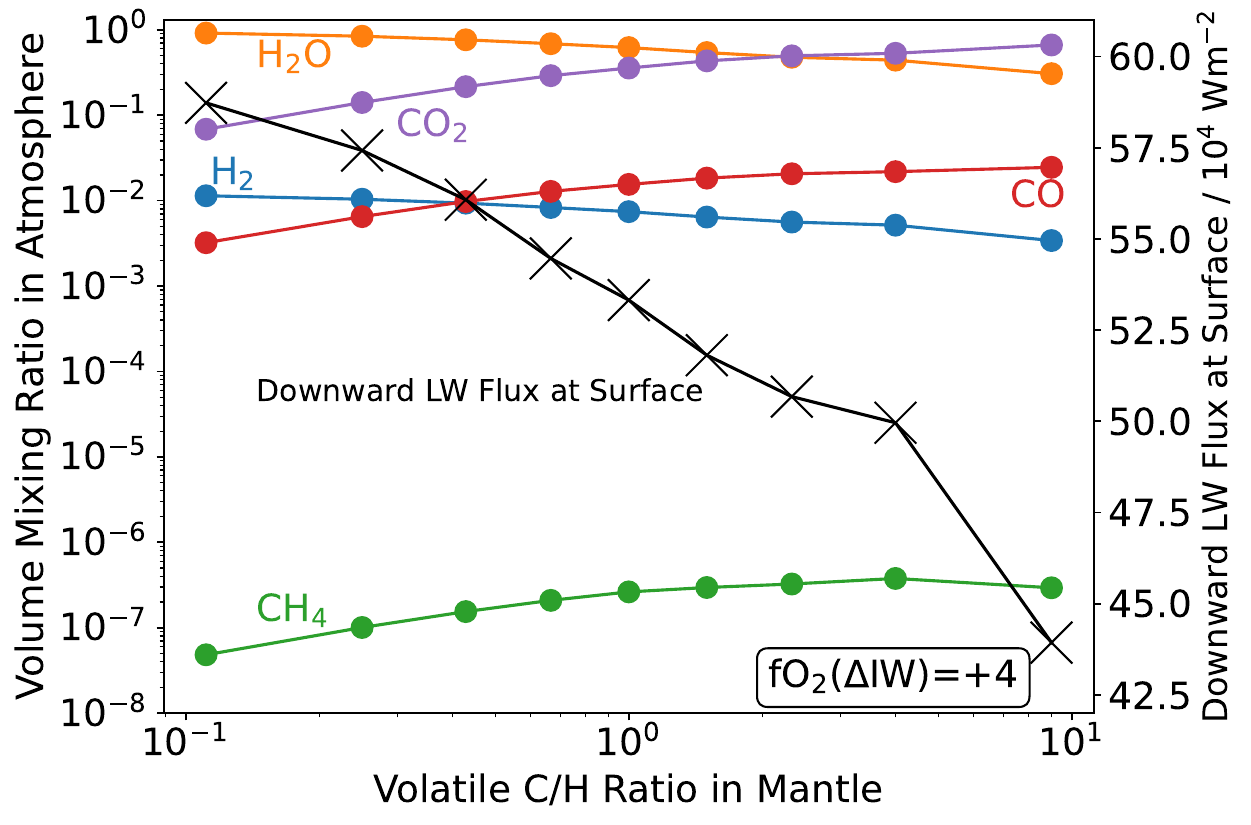}
        \label{fig:second4}
    \end{subfigure}
    \caption{Volume mixing ratios of key species in the atmosphere as a function of the C/H ratio of the bulk volatile inventory. The left and right panels show the results for simulations with mantle oxygen fugacities of $\Delta{\rm IW} = -4$ and 4 respectively. The downward longwave radiative flux at the base of the atmosphere, which quantifies the rate of surface heating due to the greenhouse effect of the atmosphere, as a function of the C/H ratio of the bulk volatile inventory is also shown.}
    \label{fig:carbonstudyspeciation}
\end{figure*}

When the volatile C/H ratio in the mantle-atmosphere system is increased for a reducing mantle ($\Delta{\rm IW} = -4$), the net atmospheric flux at the time of solidification increases with increasing carbon volatile fraction (figure \ref{fig:carbonstudyresults}). \textcolor{black}{The global melt fraction is constant for these simulations because all of the planets have solidified, i.e., the simulations terminate after reaching a global melt fraction less than 1\,\%.} This result implies that the planet is further away from the region of the parameter space corresponding to a permanent magma ocean, since the models terminate at mantle solidification, which is not necessarily at radiative equilibrium. If the C/H ratio is increased when $\Delta{\rm IW} = -4$ for the mantle, we see greater CH$_4$ volume mixing ratios (figure \ref{fig:carbonstudyspeciation}), eventually leading to CH$_4$ supplanting H$_2$ as the primary atmospheric constituent. This change in the speciation of the atmosphere leads to a smaller downward longwave radiative flux at the base of the atmosphere (figure \ref{fig:carbonstudyspeciation}), which is due to the smaller absorption cross-section of CH$_4$ relative to that due to the H$_2$-H$_2$ continuum (figure \ref{fig:molecularopacities}), thus it also leads to a reduced greenhouse heating rate of the surface.

When the volatile C/H ratio in the mantle-atmosphere system is increased for an oxidising mantle ($\Delta{\rm IW} = 4$), the global melt fraction of the planet decreases. \textcolor{black}{The net atmospheric flux is constant for these simulations because all of the planets have reached energetic equilibrium, i.e., the simulations terminate after the net flux decreases below 0.2\,W m$^{-2}$}. Similar to the results for a more reducing mantle ($\Delta{\rm IW} = -4$), an increasing C/H ratio leads to CO$_2$ replacing H$_2$O as the primary atmospheric constituent (figure \ref{fig:carbonstudyspeciation}). Again, this results in a smaller downward longwave radiative flux at the base of the atmosphere, and thus a reduced greenhouse heating rate of the surface, given the smaller absorption cross-section of CO$_2$ compared to H$_2$O over the relevant wavelength regime (figure \ref{fig:molecularopacities}). 

From these simulations, both with oxidising and reducing mantles, we can conclude that gas dwarfs with a larger intrinsic C/H ratio of the bulk volatile inventory are more likely to end their thermal evolution with a solidified mantle due to a diminished greenhouse heating rate of their surfaces. This will result in the solidification shoreline being pushed to higher instellation fluxes. However, the secondary studies undertaken so far have focused on the influence of the atmosphere on the thermal steady state of a gas dwarf. We have not accounted for the mass of the silicate-iron interior on its thermal evolution. In the final secondary parameter study, we determine the effect of the mass of the solid interior on the thermal steady state of gas dwarfs.

\subsection{Planet Mass}
\label{subsec:planetmass}

The mass and composition of the atmosphere play a key role in determining the thermal steady state of a gas dwarf, as we have shown in our primary study and our secondary parameter studies. However, the mass of the iron-silicate interior may also play a role, given that previous studies have shown its influence on the thermal evolution of gas dwarfs and partially-molten super-Earths. \cite{Herath2024} show that, for airless planets in the super-Earth regime, a larger planet mass leads to a lower global mantle melt fraction after 10 Gyrs, due to larger pressures in the interior resulting in pressure-induced solidification. The consequences of a larger or smaller silicate mass on the thermal evolution of a gas dwarf are still unclear, however. In this study, we vary the total mass of the planet and investigate its effect on the thermal steady state of gas dwarfs, and thus the location of the solidification shoreline in the $T_{\text{eff}}$-$F_{\text{ins}}$ parameter space. Note that varying the total planet mass is in effect varying the mass of the silicate interior if the EMF is held constant, given that the volatile inventory and the capacity for the mantle to store volatiles will both increase simultaneously.

We vary the total mass of the planet in our simulation between $2\,M_{\oplus}$ and $8\,M_{\oplus}$, given that masses larger than $8\,M_{\oplus}$ result in deep-mantle pressures outside the range of our MgSiO$_3$ equation of state. We increase the instellation flux to $35\,F_{\oplus}$ and the EMF to 1\%, in order to ensure that the planet is in the `permanent magma ocean' region of the parameter space. We find that the global melt fraction above the middle layer of the mantle, defined by the radial co-ordinate, is insensitive to the planet mass (figure \ref{fig:planetmassresults}). While a larger planet mass leads to pressure-induced solidification in the deep mantle \citep{Bower2018,Herath2024}, the pressures nearer the surface and thus the surface melt fraction are unaffected. Therefore, the likelihood of a gas dwarf possessing a surface magma ocean will be unaffected by its mass, implying that the location of the solidification shoreline is insensitive to planet mass. 

\textcolor{black}{One caveat to this result is that we do not account for the relationship between interior mass and EMF that arises due to atmospheric escape \citep{Rogers2021}. Therefore, our results show that the solidification shoreline is insensitive to interior mass assuming no relation between the interior mass and the EMF}. Another caveat is that we do not explore the full mass range expected for sub-Neptunes (i.e., up to and including the mass of Neptune) in this study, therefore we do not capture the full effect of the planet mass on the thermal steady state for gas dwarfs.

\begin{figure} 
    \centering
    \begin{subfigure}[b]{0.49\textwidth}
        \centering
        \includegraphics[width=\linewidth]{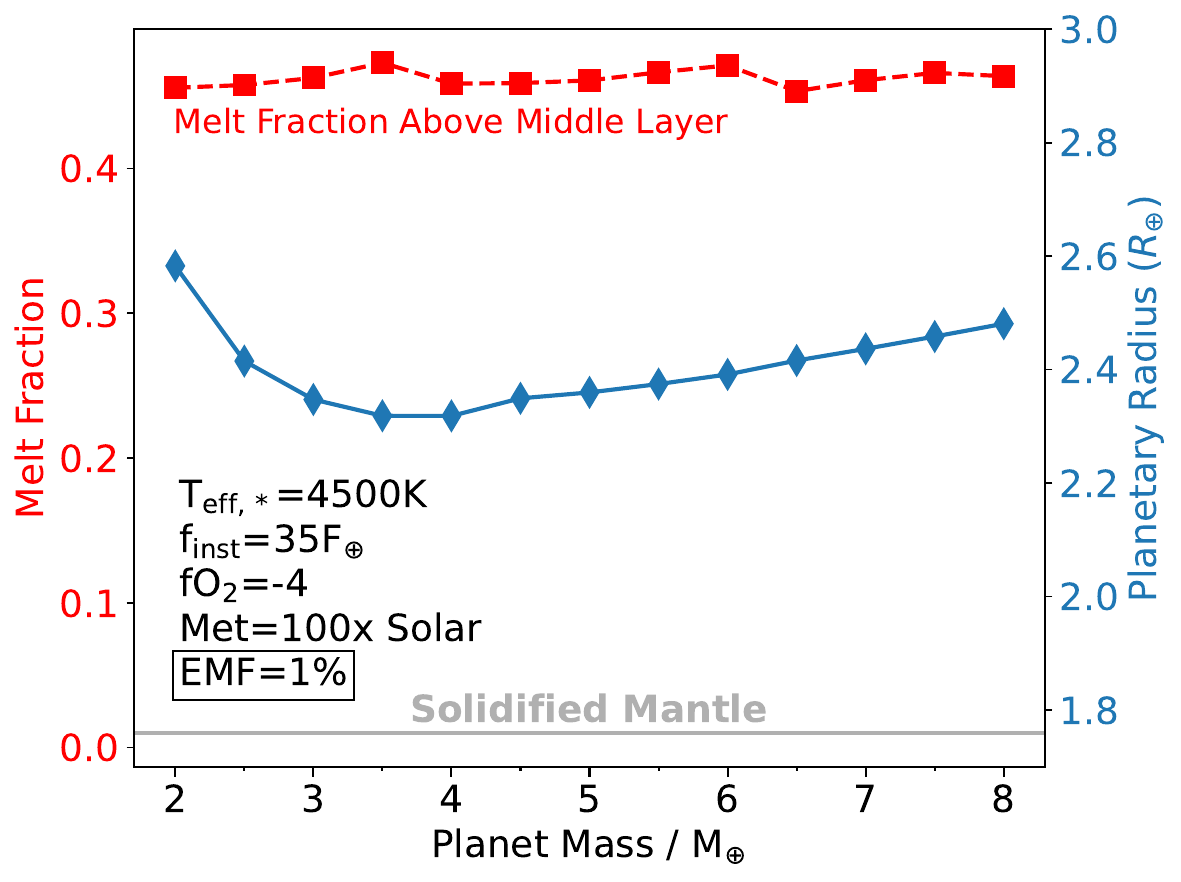}
        \label{fig:first5}
    \end{subfigure}
    \caption{Volume melt fraction in the mantle above the middle layer, defined by the radial co-ordinate, and planetary radius as a function of planet mass. The stellar effective temperature, instellation flux, oxygen fugacity, metallicity and envelope mass fraction used in these simulations are also shown.}
    \label{fig:planetmassresults}
\end{figure}

\section{Discussion}
\label{sec:discussion}

\subsection{Sensitivity of the Solidification Shoreline to Evolutionary Timescales}
\label{subsec:timescalestudy}

The division of the $T_{\text{eff}}$-$F_{\text{ins}}$ parameter space for sub-Neptunes into two regions corresponding to permanent magma oceans and solidified mantles shown in figure \ref{fig:solidificationshoreline} is only relevant with respect to the thermal steady state of gas dwarfs. The length of their cooling timescales \citep{Vazan2018,Tang2024} implies that many of the gas dwarfs in the region of the parameter space corresponding to a solidified mantle may still be cooling, and will thus still have surface magma oceans. This possibility necessitates an investigation into the timescales required to achieve either mantle solidification or a permanent magma ocean.

To study the sensitivity of the location of the solidification shoreline to the age of the planetary system, we show the time taken to reach a thermal steady state, either a permanent magma ocean or a solidified mantle, for all the simulations in this study with an EMF of 1\% (figure \ref{fig:timescalesplot}). For the planets that end their evolution with permanent magma oceans, the timescales required for them to reach a steady state are of order 1~Gyr, an order of magnitude lower than the 10~Gyr cooling timescales reported in \cite{Tang2024}. For the planets that end their evolution with solidified mantles, their solidification timescales are of order 100~Myr for $F_{\text{ins}}=0.8$--$1.4\,F_{\oplus}$ and of order 10~Myr for lower instellation fluxes: several orders of magnitude shorter than the solidification timescales presented in \cite{Tang2024}.

\begin{figure}
    \includegraphics[width=0.49\textwidth]{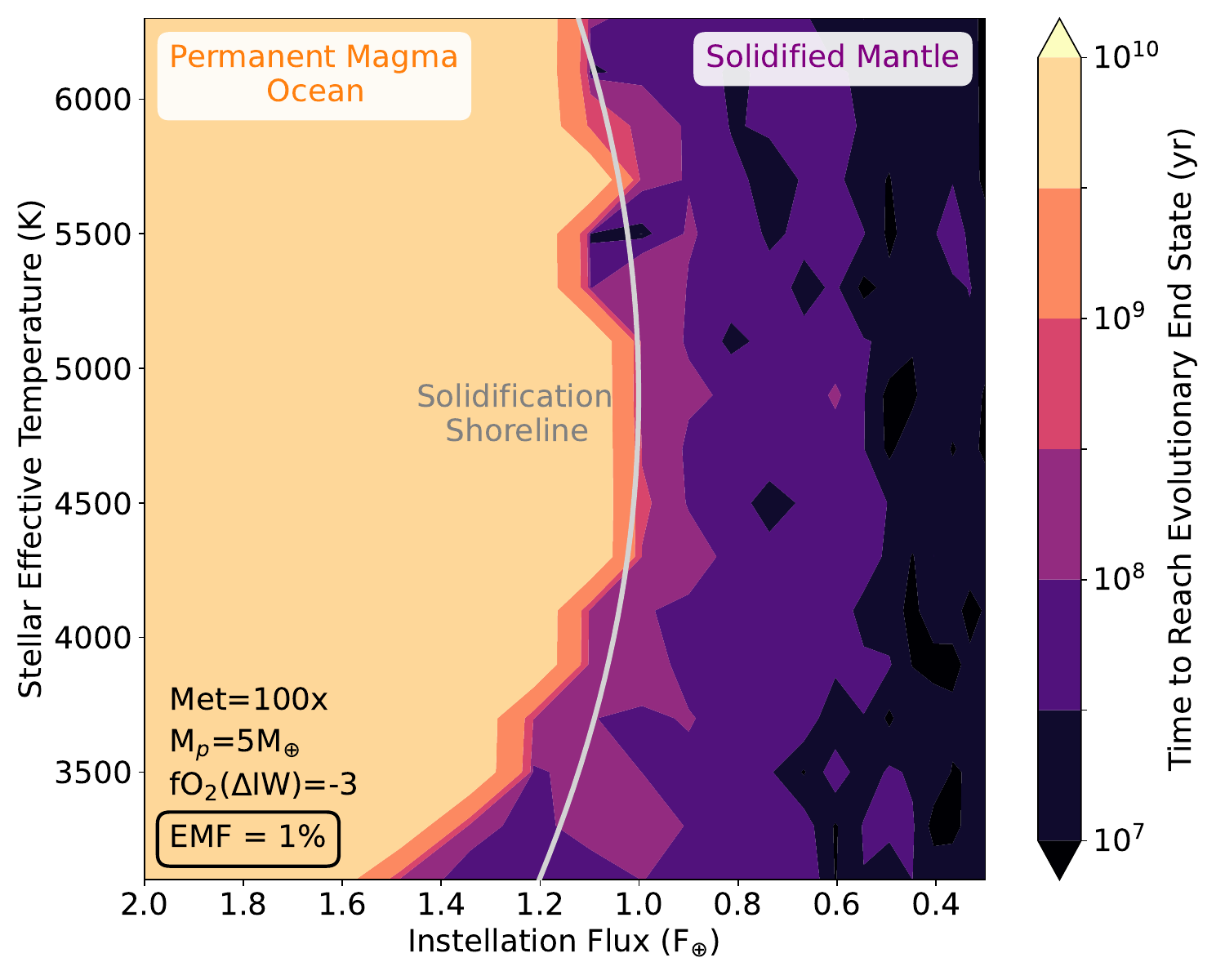}
    \caption{The time taken for gas dwarfs to achieve either a magma ocean steady state or to solidify, as a function of instellation flux and stellar effective temperature. Results are shown for planets modelled with an EMF of 1\%. The solidification shoreline is also shown as a white line. The volatile metallicity, planet mass and mantle oxygen fugacity used in these simulations is annotated.}
    \label{fig:timescalesplot}
\end{figure}

While the timescales presented in our work differ to those reported in \cite{Tang2024}, our key result --- that most detected sub-Neptunes will have magma oceans if they are gas dwarfs --- still holds regardless of the solidification timescales involved. If the gas dwarfs that are expected to solidify do so over 10$^7$--10$^8$~yrs, meaning that only the youngest planets will still be solidifying, then most detected sub-Neptunes are still expected to have permanent magma oceans according to the results of our main study. Alternatively, if the gas dwarfs that are expected to solidify do so over Gyr timescales, then there will be a significant fraction of the gas dwarf population with temporary magma oceans in addition to those with permanent magma oceans. In either scenario, the bulk of the detected gas dwarf population is still expected to be molten.

The discrepancy in the gas dwarf solidification timescales presented in this work and in \cite{Tang2024} is a consequence of several key differences in the modelling approaches used in our work and theirs. {\textcolor{black}{\cite{Tang2024} treat the envelope in their model as two distinct layers -- an upper radiative layer and a deep convection dominated layer -- whereas our atmospheric model solves for radiative-convective equilibrium, permitting deep radiative regions. Deep convectively-stable regions generate different outgoing radiation fluxes and planetary radii compared to a fully-adiabatic temperature structure, all else equal, because of their shallower lapse rate \citep{selsis_radiative_2023, cmiel_radiative_2025, Nicholls2025, Peng2024}}. Furthermore, the timescales shown in \cite{Tang2024} are for gas dwarfs with pure H-He atmospheres, whereas the timescales shown in Figure \ref{fig:timescalesplot} are for planets with atmospheres that are 10\% CH$_4$ by volume. Given the reduced surface heating that results from increasing the CH$_4$ volume mixing ratio at the expense of the H$_2$ volume mixing ratio (see section \ref{subsec:carbonbudget}), gas dwarfs with carbon-enriched atmospheres will have shorter cooling timescales compared to those with pure H-He atmospheres. Finally, the equation of state and material properties for perovskite (i.e., the solid potion of the mantle) used in our work differ from those used in \cite{Tang2024}. This will result in different rates of heat transport through the solid material and thus different solidification timescales between the two models. 

Regardless of the insensitivity of our conclusions to solidification timescales, it is important to understand the sensitivity of these evolutionary timescales to planetary properties, as their length could affect the atmospheric composition of gas dwarfs, if not the likelihood of them having a permanent magma ocean. Also, if solidification timescales are of order 10$^7$-10$^8$ yrs, this motivates observational campaigns targeted at young sub-Neptunes \citep{Barat2025AJ} to study the transition from a magma oceans to a solidified mantle in the sub-Neptune regime. Further work to determine the sensitivity of the location of the solidification shoreline to cooling timescales would help resolve the discrepancy between our solidification timescales and those of \cite{Tang2024}, and assist in estimating the melt state of young sub-Neptunes.

\subsection{Consistency of Simulated Planets With the Gas Dwarf Hypothesis}
\label{subsec:mmwconstraints}

{\color{black}

The simulated planets presented in this work span a wide region of the parameter space defined by the stellar effective temperature, instellation flux, EMF, oxygen fugacity, volatile C/H ratio and planet mass. Here, we return to a key question: are these planets still gas dwarfs, i.e., do they fulfil the criteria outlined in table \ref{tab:gasdwarfwaterworldcriteria}? 

One criterion that many of our simulated planets are unlikely to fulfil is the upper limit we impose on the mean molecular weight. For the studies in which we vary the oxygen fugacity of the mantle (figure \ref{fig:oxygenfugacity}) and the volatile C/H mass ratio of the mantle (figure \ref{fig:carbonstudyresults}), the compositions of the simulated atmospheres deviate substantially from the hydrogen-dominated atmospheres that we associate with gas dwarfs. As a consequence, the mean molecular weights of these atmospheres increase above the upper limit we impose as a part of the gas dwarf classification. Specifically, mantle oxygen fugacities greater than $-$2 in log units relative to the iron-w\"ustite buffer and volatile C/H mass ratios greater than 0.67 exceed this upper limit (figure \ref{fig:mmwfigure}). This implies that the oxygen fugacity and the volatile C/H ratio of the mantle can only vary within a narrow parameter space while remaining consistent with our gas dwarf classification. Therefore, the solidification shoreline, applied specifically to gas dwarfs, is inherently insensitive to the oxygen fugacity and the volatile C/H ratio of the mantle.
}

\begin{figure*}
    \includegraphics[width=\textwidth]{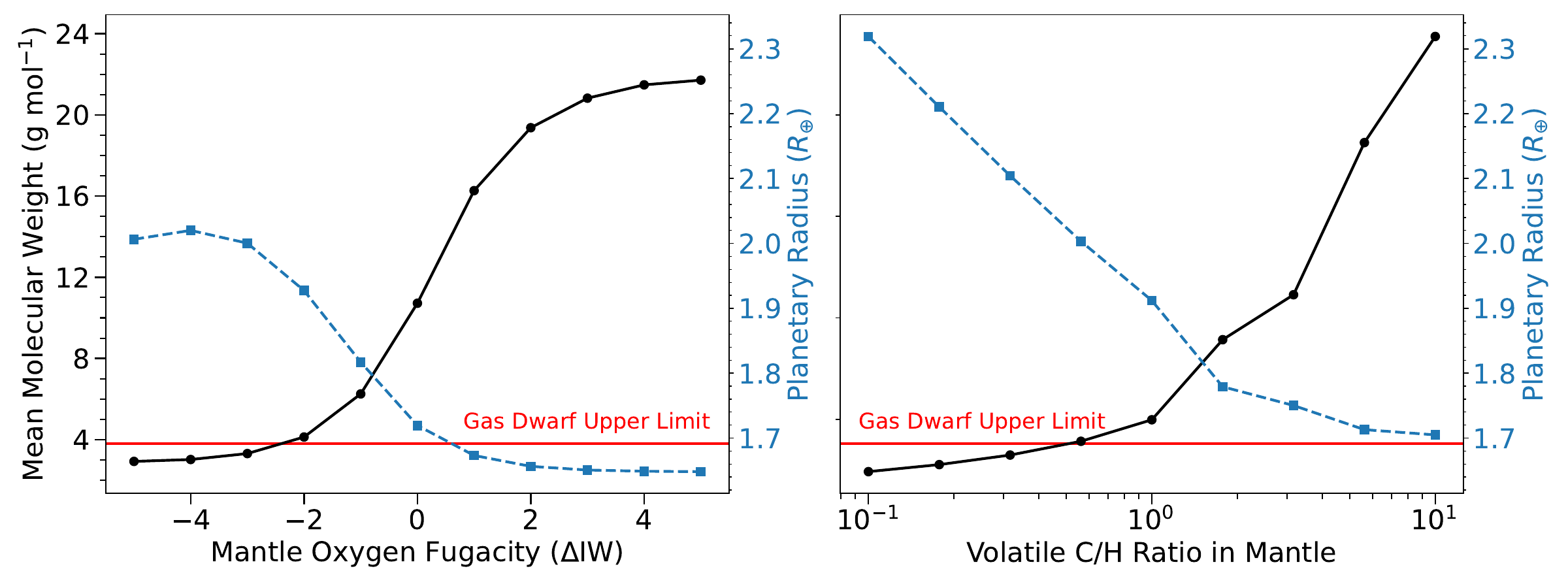}
    \caption{Atmospheric mean molecular weight and planetary radius as a function of oxygen fugacity (left) and volatile C/H mass ratio of the mantle (right). All simulations use the fiducial parameter values specified in table \ref{tab:constantparameters}. The simulations in which the volatile C/H ratio is varied are assigned a mantle oxygen fugacity of -4 in log units relative to the iron wu\"stite buffer. The red line shows the upper limit for the mean molecular weight we impose for the gas dwarf classification (table \ref{tab:gasdwarfwaterworldcriteria}).}
    \label{fig:mmwfigure}
\end{figure*}

\subsection{Sub-Neptunes With Surface Pressures Too Large to Sustain Magma Oceans}
\label{subsec:gasdwarflargeEMF}

One of the main findings in this work is that gas dwarfs with higher envelope mass fractions are more likely to end their thermal evolution with a permanent magma ocean, due to a larger greenhouse heating rate of the surface. However, if the envelope mass fraction increases to a sufficiently high value, the surface pressure will increase to the point where the surface mantle could transition from a molten state to a solid state. \cite{Breza2025} show that, for sub-Neptunes as massive as 16\,M$_{\oplus}$ with an atmospheric mean molecular weight of 10 g/mol, an EMF larger than 10\% can result in surface pressures large enough to induce a liquid to solid phase change. This implies that the sub-Neptunes shown in figure \ref{fig:solidificationshoreline}, if their envelopes have a high mean molecular weight and they have an EMF greater than 10\%, may be found in the solidified region of the parameter space.

\textcolor{black}{While the existence of such planets is a caveat to our findings, modelling them is outside the scope of this work. According to our criteria, these planets would not be classified as gas dwarfs, given that their mean molecular weights exceed 3.8\,g mol$^{-1}$. Nevertheless, the presence and thermal evolution of magma oceans on sub-Neptunes with high mean molecular weight atmospheres is a pertinent research area that requires detailed investigation, given that outgassing thermodynamics can result in sub-Neptune atmospheres with mean molecular weights ranging from 2 to 10\,g mol$^{-1}$ \citep{Heng2025}. Such investigation could involve the extension of the solidification shoreline to sub-Neptunes outside of the gas dwarf classification, as we have defined it.}

\subsection{The Impact of Atmospheric Escape on the Solidification Shoreline}
\label{subsec:escapejustifiaction}

In our modelling framework, we do not account for atmospheric escape through either photo-evaporative or core-powered mass loss. These processes are predicted to significantly erode the atmospheres of close-in sub-Neptunes \citep{Mordasini2020,Rogers2023,Tang2024}, reducing the atmospheric mass fractions over the course of their thermal evolution. We have shown that the location of the solidification shoreline is strongly dependent on the atmospheric mass fraction. Therefore, the decrease of the atmospheric mass fraction could reduce the greenhouse heating rate of the surface, which could affect the thermal steady state of a gas dwarf. This would appear to suggest that the lack of atmospheric escape within our modelling framework is a major caveat to our results.

However, while atmospheric escape will affect the thermal evolution of a gas dwarf, it will not affect our finding that most observed sub-Neptunes, if they are gas dwarfs, will be molten. \textcolor{black}{Equation 2 in \cite{Lopez2014}, as used in this work,} predicts the EMF that a sub-Neptune would have currently if it were a gas dwarf, based on its observed radius. If the planet in question were susceptible to atmospheric escape, then its EMF would have been greater earlier in its evolutionary history. The increased EMF would have made it even more likely for the planet to end its thermal evolution with a permanent magma ocean, given the additional surface heating induced by a more massive atmosphere. Similarly, the total volatile inventories in our simulations can be interpreted as the minimum volatile inventory, and thus atmospheric mass fraction, that the simulated planet would have over the course of its evolution. If a real planet had a larger volatile inventory in the past, which had been eroded by atmospheric escape, then the planet would be more likely to have a permanent magma ocean today. Therefore, the solidification shorelines presented in figure \ref{fig:solidificationshoreline} can still be used to predict thermal steady states for the detected sub-Neptune population.

\subsection{Implications for Sub-Neptunes with Observed JWST Transmission Spectra}
\label{subsec:observedsubneptunes}

{\color{black}

Given that 98\% of the observed sub-Neptune population lies within the `permanent magma ocean' region of the $T_{\text{eff}}$-$F_{\text{ins}}$ parameter space, it is germane to ask whether or not observations of sub-Neptunes within this region are consistent with their being gas dwarfs with magma oceans. In figure \ref{fig:solidificationshoreline}, we show the locations of several sub-Neptunes with published JWST transmission spectra. We consider the observational constraints on the masses, radii and atmospheric mean molecular weights of a subset of these planets in the context of our results, and assess their consistency with the molten gas dwarf hypothesis.

Several of these planets, while within the `permanent magma ocean' region of the parameter space in our study, are disfavoured as gas dwarfs. JWST transmission spectra for GJ 9827 d \citep{Piaulet2024} and GJ 1214 b \citep{Ohno2025} both favour high mean molecular weight atmospheres for their respective planets. While their surfaces may or may not be molten \citep[cf.][]{Breza2025}, we do not classify them as gas dwarfs on the basis of these high mean molecular weights; therefore, our results are not relevant for these planets. Observations of TOI-836 c, however, are less conclusive about its mean molecular weight and atmospheric chemistry \citep{Wallack2024}. The mass and radius constraints for the planet are consistent with the mass-radius regime that we have identified for molten gas dwarfs (figure \ref{fig:massradiusplot}), and they lie within the `permanent magma ocean' region of the $T_{\text{eff}}$-$F_{\text{ins}}$ parameter space (figure \ref{fig:solidificationshoreline}). In light of this finding, future observations of TOI-836 c could be interpreted in the context of a gas dwarf with a magma ocean, alongside other interior structures.

}

\begin{figure}
    \includegraphics[width=0.49\textwidth]{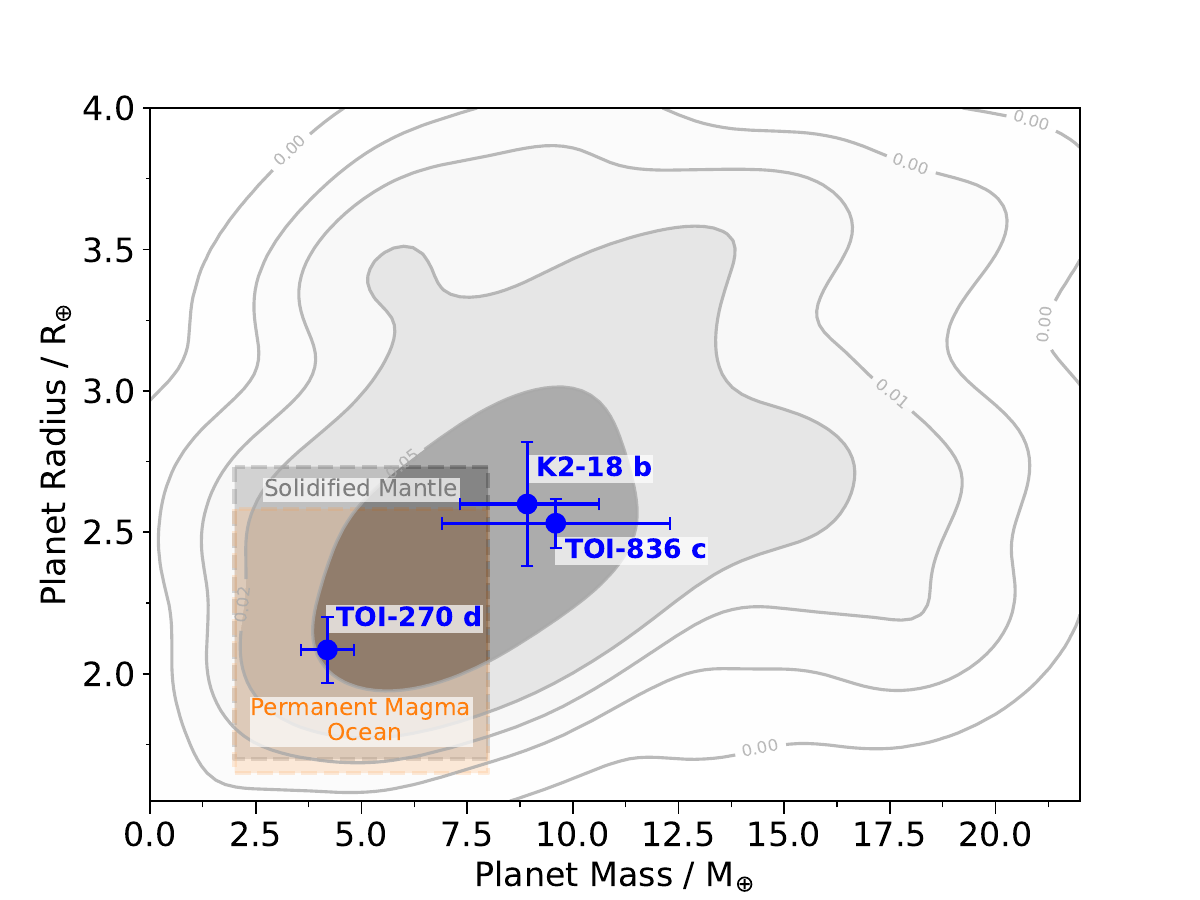}
    \caption{Rectangular boxes showing the range of planetary masses and radii for molten gas dwarfs (orange region) and the solidified gas dwarfs (grey region) modelled in this work. A kernel density contour plot for the mass-radius distribution for all detected sub-Neptunes is also shown. K2-18 b, TOI-270 d and TOI-836 c are also plotted in blue.}
    \label{fig:massradiusplot}
\end{figure}

Different observations of TOI-270 d with transmission spectroscopy have been used to infer different conclusions regarding the mean-molecular weight of its atmosphere. \cite{Benneke2024} report a measurement of $\mu$=5.47$^{+1.25}_{-1.14}$, which favours a water-rich, mini-Neptune scenario. However, this result may still be consistent with a molten interior and miscible atmosphere \citep{Glein2025}. Conversely, \cite{Holmberg2024} argue that the transmission spectrum reported in their work is consistent with a H$_2$-dominated, low mean molecular weight atmosphere. The mass and radius constraints for TOI-270 d are consistent with a molten gas dwarf (figure \ref{fig:massradiusplot}), and our results show that TOI-270 d would have a permanent magma ocean if it were a gas dwarf (figure \ref{fig:solidificationshoreline}). Also, retrievals performed on the transmission spectrum of JWST presented in \cite{Holmberg2024} show evidence of CH$_4$ and CO$_2$ in the upper atmosphere of TOI-270 d as well as a non-detection of NH$_3$, which could be explained by the presence of a magma ocean \citep{Shorttle2024,Rigby2024}. Lastly, \cite{Nixon2025} demonstrate that chemical interactions between a magma ocean and an H$_2$-rich atmosphere can explain the observational constraints for H$_2$O and CO$_2$ for TOI-270 d in \cite{Benneke2024}. This leaves the molten gas dwarf hypothesis as a viable explanation for the observations of TOI-270 d.

Finally, we consider the observations of the sub-Neptune K2-18 b in the context of our findings. K2-18 b has been the subject of considerable attention from an observational perspective \citep{Cloutier2017,Sarkis2018,Benneke2019,Madhusudhan2023,Madhusudhan2025,Hu2025} and a theoretical perspective \citep{Wogan2024,Cooke2024,Shorttle2024,Rigby2024,Gupta2025,Liu2025}. Figures \ref{fig:solidificationshoreline} and \ref{fig:massradiusplot} show that the mass, radius and instellation flux constraints for K2-18 b are consistent with a gas dwarf with a permanent magma ocean. Therefore, our results cannot rule out the molten gas dwarf hypothesis for K2-18~b using these constraints. To further explore this hypothesis for K2-18 b, we must compare self-consistent models of it as a gas dwarf to its observed transmission spectrum. While this has been done previously \citep{Shorttle2024,Rigby2024}, an evolutionary modelling framework has yet to be used to explore the molten gas dwarf hypothesis for K2-18 b.

An important caveat to our prediction that K2-18 b and TOI-270 d are expected to have permanent magma oceans in the gas dwarf paradigm is that our estimation of the location of the solidification shoreline may be affected by our use of blackbody stellar spectra. Our sensitivity tests comparing the PT profiles generated by our climate model using synthetic stellar spectra and blackbody spectra (section \ref{apsec:blackbodytests}) imply that the solidification shoreline for planets orbiting M-type stars may be located at higher instellation fluxes than suggested by figure \ref{fig:solidificationshoreline}. Given that both of these planets orbit M-type stars, they may be located closer to the boundary between solidification and a permanent magma ocean in the $T_{\text{eff}}$-$F_{\text{ins}}$ parameter space.

\subsection{Future Development of the Solidification Shoreline}
\label{subsec:modellimitations}

While we have presented the solidification shoreline as a metric for evaluating whether sub-Neptunes are expected to be molten or solid, there is further work to done in developing this metric. An improvement to our modelling framework would be the inclusion of volatile storage in the solid-phase of the mantle as well as the metallic core, through partitioning into the solid and/or trapping of melt \citep{tikoo_fate_2017,Hier2017,Sim2024}. Incorporating processes such as stellar evolution and orbital migration into an evolutionary modelling framework such as \texttt{PROTEUS} could improve the accuracy of our estimation of the location of the solidification shoreline, given that these processes determine the instellation flux received by the planet. Exploring the full range of planet masses relevant for sub-Neptunes would also refine the robustness of our prediction that most observed gas dwarfs have permanent magma oceans. Using real or synthetic stellar spectra as opposed to blackbody spectra would help improve the accuracy of our estimate of the location of the solidification shoreline, especially for M-type stars.

Finally, there is the potential for extending the use of the solidification shoreline as a metric for all sub-Neptunes, rather than for gas dwarfs only. This would enable evaluation of the magma ocean hypothesis for sub-Neptunes with high mean molecular weight and/or miscible atmospheres. Such a metric would involve major improvements to our modelling framework, including solubility laws that are valid for the surface pressures expected for these planets and possibly even volatile-silicate miscibility \citep{Rogers2025b}. It would also involve accounting for the possibility of large surface pressures preventing the existence of magma oceans within an evolutionary modelling framework \citep{Breza2025}. 

\section{Conclusions}
\label{sec:conclusions}

 The degeneracy in the interior structures of sub-Neptunes necessitates the use of spectroscopy to obtain information about their atmospheric chemistry that can be used to break this degeneracy. If gas dwarfs, sub-Neptunes with earth-like interiors and low mean-molecular weight atmospheres, have surface magma oceans, then the chemical interaction between their magma oceans and their atmospheres could produce chemical signatures that are spectroscopically identifiable. However, it is unclear how common magma oceans are on gas dwarfs. We address this gap by using a 1D coupled climate-interior evolution model to identify the location of the `solidification shoreline': the boundary in instellation flux vs. stellar effective temperature parameter space that separates solidified gas dwarfs from those which maintain permanent magma oceans at steady-state. Our conclusions are as follows:

\begin{itemize}
    \item 98\% of detected sub-Neptunes are in the region of the $T_{\text{eff}}$-$F_{\text{ins}}$ parameter space corresponding to permanent, steady state magma oceans for gas dwarfs. Therefore, most observed sub-Neptunes, if they are gas dwarfs, have permanent magma oceans. If a sizeable fraction of the sub-Neptune population are gas dwarfs, this motivates theoretical work to understand and predict the consequences of interactions between the magma and the atmospheres of these planets, as well as dedicated observational campaigns to search for chemical signatures of these interactions.
    \item The most significant parameters that determine the steady state melt fraction of gas dwarfs are the instellation flux and the atmospheric mass fraction. The stellar effective temperature exerts negligible influence on their thermal steady state when the instellation flux is kept constant.
    \item The C/H ratio of the bulk volatile inventory and the mantle oxygen fugacity exert some influence over the thermal steady state, with higher volatile C/H ratios and more reducing mantles making a solidified mantle a more likely evolutionary outcome. However, planets with higher volatile C/H ratios and more oxidising mantles also \textcolor{black}{have high atmospheric mean molecular weights and are thus outside the scope of this study.} While their degree of deep-mantle melting is sensitive to planet mass, the degree of melting near the surface is not; for this reason, the likelihood of gas dwarfs possessing a surface magma ocean will be unaffected by their mass.
    \item Of the sub-Neptunes for which there are JWST transmission spectra, we find that three planets (TOI-836 c, TOI-270 d, K2-18 b) have masses, radii and instellation fluxes consistent with them being gas dwarfs with permanent magma oceans, although this prediction is sensitive to the choice of host stellar spectrum used. Given that low mean molecular weight atmospheres have not been ruled out for these planets, a molten gas dwarf scenario cannot be ruled out for these planets based on their masses, radii and instellation fluxes alone. Future observational constraints on their atmospheric chemistry, combined with self-consistent evolutionary modelling frameworks, are required to further explore the magma ocean hypothesis for these planets.
\end{itemize}

\section*{Acknowledgments}

R.C. thanks the Science and Technology Facilities Council (STFC) for the PhD studentship (grant reference ST/Y509139/1). O.S. and H.N. acknowledge support from STFC grant UKRI1184. C.M.G. was supported by the STFC (grant reference ST/W000903/1) and by an ETH Postdoctoral Fellowship. T.L. was supported by the Branco Weiss Foundation, the Netherlands eScience Center (PROTEUS, NLESC.OEC.2023.017), the Alfred P. Sloan Foundation (AEThER, G202114194), NASA’s Nexus for Exoplanet System Science research coordination network (Alien Earths, 80NSSC21K0593), and the Dutch Research Council NWA-ORC PRELIFE Consortium (NWA.1630.23.013).

\section*{Data Availability}

The data underlying this article will be shared on reasonable request to the corresponding author. The modelling framework used in this work is fully open-source; there are GitHub pages for \texttt{PROTEUS} (https://github.com/FormingWorlds/PROTEUS), \texttt{AGNI} (https://github.com/nichollsh/AGNI), \texttt{SPIDER} (https://github.com/FormingWorlds/spider), \texttt{SOCRATES} (https://github.com/nichollsh/SOCRATES) and \texttt{CALLIOPE} (https://github.com/FormingWorlds/CALLIOPE).


\bibliographystyle{mnras}
\bibliography{references.bib} 



\appendix
\label{app:mainappendix}

\section{Solubility Laws}
\label{appsec:solubilitylaws}

Given the large pressures expected at the interfaces between sub-Neptune interiors and atmospheres \citep{Bower2025}, it is important to state the pressure-temperature conditions for which solubility laws have been experimentally determined. In this section, we outline all of the solubility laws used in our work. All pressures and fugacities are given in bar and concentrations are given in ppm by weight. Units for coefficients are omitted for brevity. \textcolor{black}{The simulations in this work span surface temperatures ranging from 1633\,K to 3820\,K and surface pressures ranging from 0.7\,GPa to 9\,GPa.}

We adopt the H$_2$O solubility law for molten peridotite from \cite{SOSSI2023}, determined experimentally at a temperature and pressure of 2173$\pm50$K and 1 bar respectively:

\begin{equation}
    X_{\mathrm{H_2O}} = 525 \times \left( p_{\mathrm{H_2O}} \right)^{0.5}.
\end{equation}

We adopt the CO$_2$ solubility law in alkalic basalt from \cite{Dixon1997}, determined experimentally at temperatures of 1673K and for pressures up to 20 kbar:

\begin{equation}
    X_{\mathrm{CO_3}^-} = 3.8 \times 10^{-7}\ \times p \exp\!\left( \frac{-23(p-1)}{83.15\,T} \right),
\end{equation}

\begin{equation}
    X_{\mathrm{CO_2}} = 10^{4} \times \frac{4400\,X_{\mathrm{CO_3}^-}}{36.6 - 44\,X_{\mathrm{CO_3}}}.
\end{equation}

We adopt the solubility law for N$_2$ in basalt from \cite{DASGUPTA2022}, taking X$_{\text{SiO}_2}$=0.56, X$_{\text{Al}_2\text{O}_3}$=0.11, X$_{\text{TiO}_2}$=0.01. This law is valid up to pressures of 9GPa and temperatures of 1600K:

\begin{equation}
    A = \exp\!\left[ 4.67 + 7.11 X_{\mathrm{SiO_2}} - 13.06 X_{\mathrm{Al_2O_3}} - 120.67 X_{\mathrm{TiO_2}} \right],
\end{equation}

\begin{equation}
    X_{\mathrm{N_2}} =
\left( \frac{p_{\mathrm{N_2}}}{10000} \right)^{0.5}
\exp\!\left[
\frac{5908.0 \times (p/10000)^{0.5}}{T}
- 1.6\,\Delta \mathrm{IW}
\right]
+ \frac{A\,p_{\mathrm{N_2}}}{10000},
\end{equation} where $\Delta \mathrm{IW}$ refers to the oxygen fugacity relative to the iron-w\"ustite buffer in log units.

We adopt the solubility law for CH$_4$ in basalt from \cite{Ardia2013}, valid for pressures ranging from 0.7GPa to 3GPa and temperatures ranging from 1673K to 1723K:

\begin{equation}
    X_{\mathrm{CH_4}} =
\left( \frac{p_{\mathrm{CH_4}}}{10000} \right) \times
\exp\!\left[ 4.93 - 1.93 \times 10^{-8} p \right].
\end{equation}

We adopt the solubility law for CO in basalt from \cite{Armstrong2015}, based on experiments conducted at 1.2GPa and 1673K:

\begin{equation}
\log_{10} X_{\mathrm{CO}}
= -0.738 + 0.876 \log_{10}\!\left( p_{\mathrm{CO}} \right)
- 5.44 \times 10^{-5} p.
\end{equation}

\section{Atmospheric Mass in Secondary Parameter Studies}
\label{apsec:atmosmasssecondstudies}

\begin{figure}
    \includegraphics[width=0.49\textwidth]{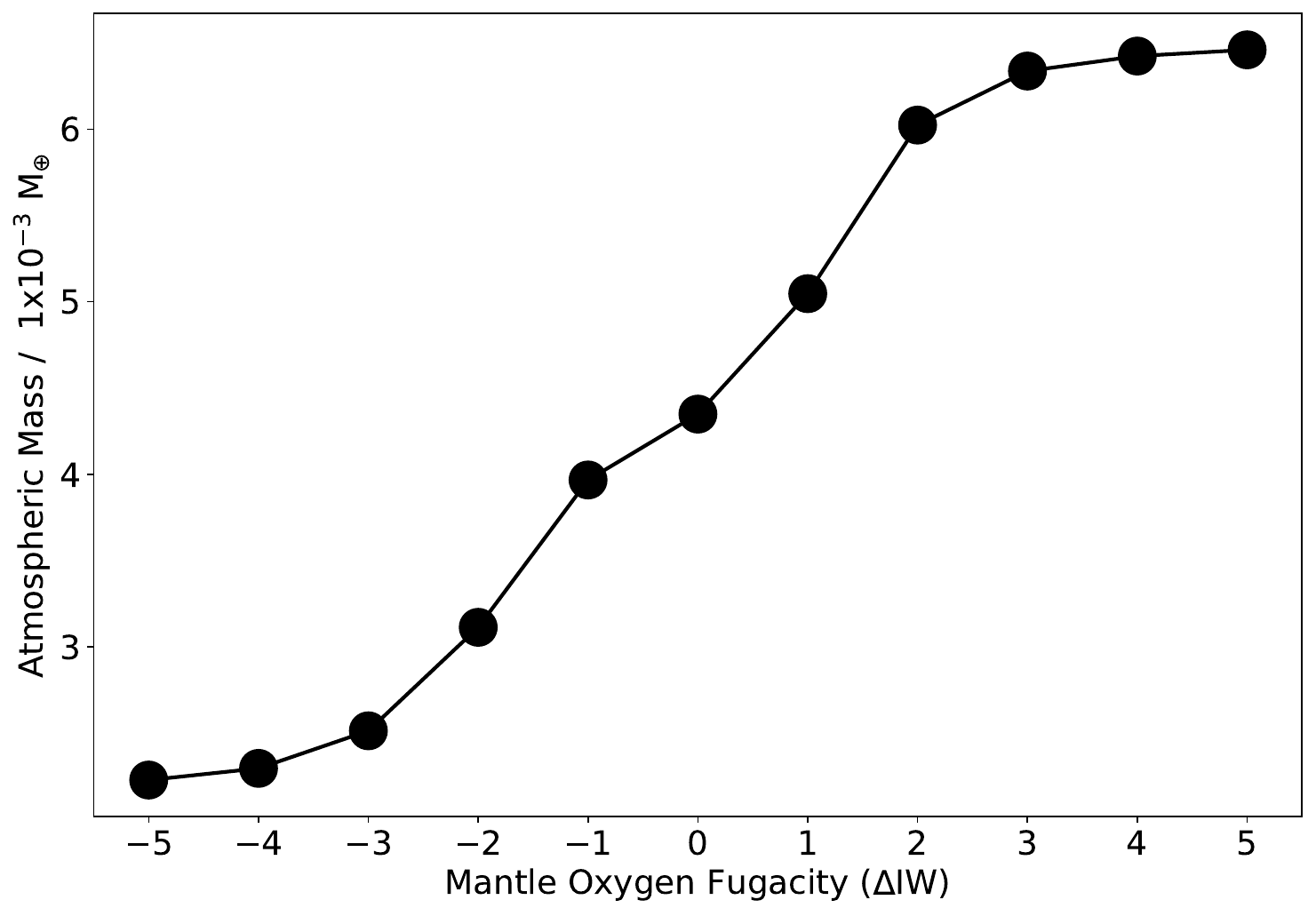}
    \caption{Atmospheric mass as a function of oxygen fugacity ($\Delta$IW) of the mantle for the simulations in which we vary the oxygen fugacity of the magma ocean. The increase in atmospheric mass is due to the transition from an H$_2$-dominated atmosphere to an H$_2$O dominated atmosphere, given the larger mean molecular weight of H$_2$O compared to H$_2$.}
    \label{fig:oxygenfugacityatmmasses}
\end{figure}

\begin{figure}
    \includegraphics[width=0.49\textwidth]{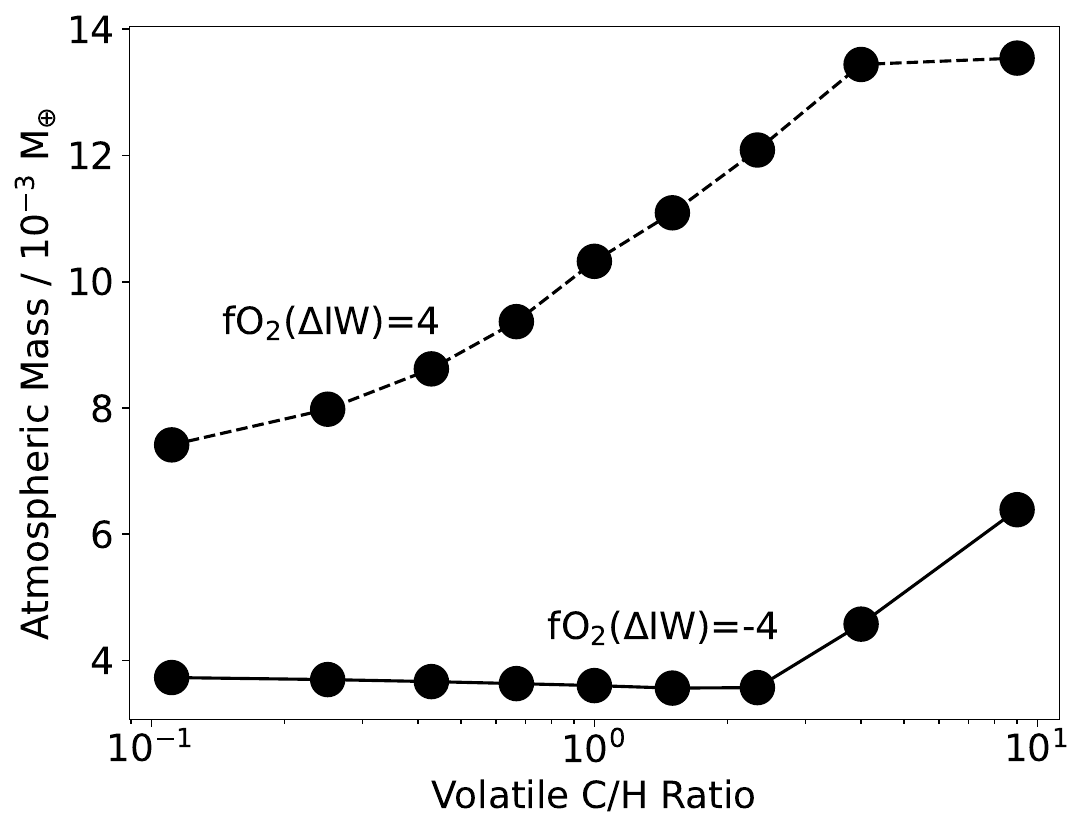}
    \caption{Atmospheric mass as a function of the C/H ratio of the bulk volatile inventory Results are shown for simulations with a mantle oxygen fugacity ($\Delta$IW) of -4 and 4 as solid and dashed lines respectively.}
    \label{fig:carbonstudyatmmasses}
\end{figure}

\section{Blackbody Spectra Sensitivity Tests}
\label{apsec:blackbodytests}

We conduct sensitivity tests to determine how the pressure-temperature profiles generated by our 1D radiative-convective climate model (\texttt{AGNI}) differ when blackbody spectra are used to compute radiative fluxes as opposed to real or simulated stellar spectra. These tests were conducted using G, K and M type solar spectra.  We run \texttt{AGNI} with the solar spectrum from \cite{Gueymard2004} and synthetic spectra for HD85512 (T$_{\text{eff}}$=4404K) and GJ849 (T$_{\text{eff}}$=3467K) from the MUSCLES survey \citep{France2016}. We also run \texttt{AGNI} using three blackbody spectra, each with blackbody temperatures of 5772K, 4404K and 3467K, to represent blackbody approximations of the spectra of the Sun, HD85512 and GJ849. We run \text{AGNI} using input parameters representative of a typical sub-Neptune in our study (see table \ref{tab:bbtestinputs}).

\begin{table}
\centering
\begin{tabular}{l|l}
Sensitivity Test Input Parameter & Value \\
\hline
Surface Temperature & 1000\,K \\
Instellation & 1368\,Wm$^{-2}$ \\
Instellation scale factor & 0.375 \\
Characteristic solar zenith Angle & 48.19\,$^\circ$ \\
Planet Radius & 2.6\,R$_{\oplus}$ \\
P$_{\text{surf}}$ & 10$^3$\,bar \\
P$_{\text{top}}$ & 10$^{-5}$\,bar \\
H$_2$ VMR & 0.92 \\
H$_2$O VMR & 0.0097 \\
CO$_2$ VMR & 0.00555 \\
CH$_4$ VMR & 0.06 \\
\end{tabular}
\caption{Input parameters used for 1D radiative-convective climate code (\texttt{AGNI}) when conducting tests comparing PT profiles calculated using blackbody spectra as opposed to real or synthetic stellar spectra.}
\label{tab:bbtestinputs}
\end{table}

\begin{figure}
    \includegraphics[width=0.49\textwidth]{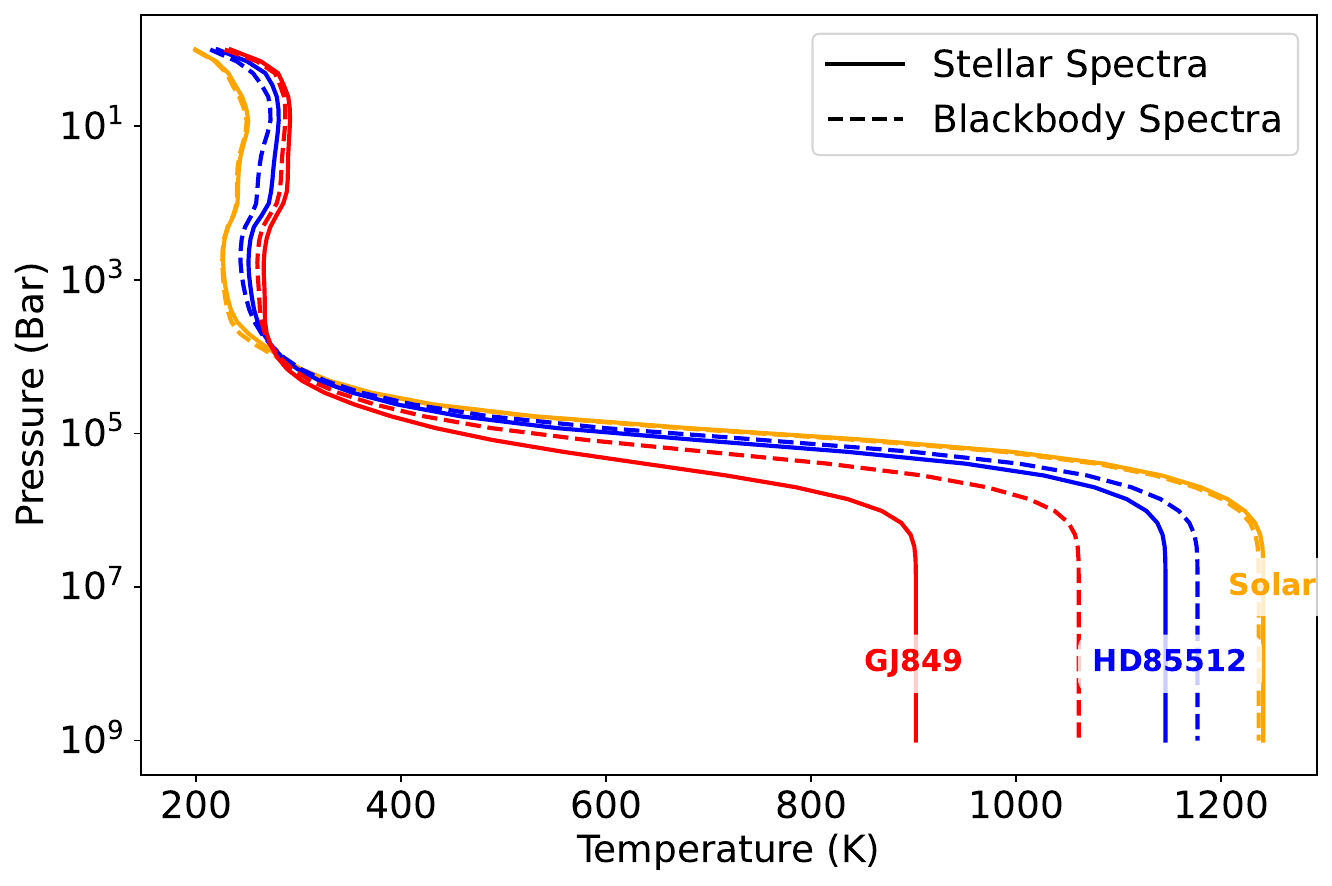}
    \caption{Pressure-temperature (PT) profiles calculated using our radiative-convective climate code (\texttt{AGNI}). Solid lines correspond to a PT profile calculated using a real or synthetic spectra, and dashed lines correspond to a profile calculated using a blackbody spectra. Profiles corresponding to a solar spectra \citep{Gueymard2004} or a blackbody equivalent (T$_{\text{eff}}$=5772\,K) are shown in yellow. Profiles corresponding to MUSCLES spectra \citep{France2016} of HD85512 and GJ849, or their blackbody equivalents (T$_{\text{eff}}$=4404\,K and 3467\,K respectively) are shown in blue and red respectively.}
    \label{fig:bbcomparisonfigure}
\end{figure}

For the solar spectrum, the temperature difference in the PT profile calculated using a real spectrum and a blackbody equivalent is negligible (figure \ref{fig:bbcomparisonfigure}). For a K-type stellar spectrum such as that of HD85512, the temperature difference between the synthetic stellar spectrum and the blackbody near the surface is small: \textasciitilde40\,K. However, for the spectrum of GJ849, the difference in surface temperature calculated using the synthetic stellar spectrum is \textasciitilde200\,K colder than that calculated using the blackbody spectrum. This suggests that we are underestimating the surface temperatures of gas dwarfs orbiting M-type stars, implying that the solidification shoreline may be located at lower instellation fluxes than those suggested by figure \ref{fig:solidificationshoreline}.

\section{Sulfur Species Sensitivity Tests}
\label{apsec:sulfurtests}

{\color{black}

This work focusses on gas dwarf sub-Neptunes, which have low metallicity atmospheric compositions (Table~\ref{tab:gasdwarfwaterworldcriteria}). Thus, we can neglect sulfur species from our main grid of simulations because their abundance within low molecular weight atmospheres is necessarily limited. However, the thermochemical formation of H$_2$S within reducing H$_2$-dominated environments could be favourable, given sufficient sulfur content, potentially introducing an additional source of opacity alongside other volatiles \citep{Jordan2025b, Nicholls2025c, desch_the_2020}.

Here, we use our radiative-convective atmosphere model (Section~\ref{sec:methods}) to probe the potential influence of H$_2$S -- the primary carrier of sulfur -- on the climate of hydrogen rich sub-Neptunes, by comparison against H$_2$O. We consider a range of H$_2$S and H$_2$O volume mixing ratios in an H$_2$ background, for a K2-18\,b sized planet exposed to 10 times Earth's solar instellation flux, adopting zero internal heat production and $p_s=1$\,kbar.

The resultant temperature profiles plotted in Figure~\ref{fig:sulfur} panels (a) and (c) show that addition of H$_2$O and H$_2$S acts to increase the deep-atmospheric temperature, all else equal. Across the subset of these cases consistent with our gas-dwarf definition, the surface temperature increase caused by the H$_2$S greenhouse forcing is $\sim100$\,K. Importantly, comparison of panels (a) and (d) shows that, because H$_2$S is a weaker greenhouse gas than H$_2$O, it has a smaller impact on sub-Neptune climate structure than H$_2$O \citep{azzam_exomol_2016}. These behaviours validate our choice to neglect H$_2$S from the main simulations (Section~\ref{sec:solidshoreline}), since other volatile gases (e.g. water, which is included) exert more control over corresponding melting states.

In addition to temperature profiles, Figure~\ref{fig:sulfur} panels (e) and (f) plot the corresponding outgoing emission spectra for these H$_2$S-enhanced scenarios. Multiple absorption features are generated by H$_2$S, which explains the temperature increase shown by panels (c) and (d). Our main analysis does not specifically consider the spectroscopic effects of composition, however, careful consideration of these absorption features would be important for identifying the particular nature of specific sub-Neptune exoplanets \citep{Nicholls2025c} 
}

\begin{figure}
    \includegraphics[width=\linewidth]{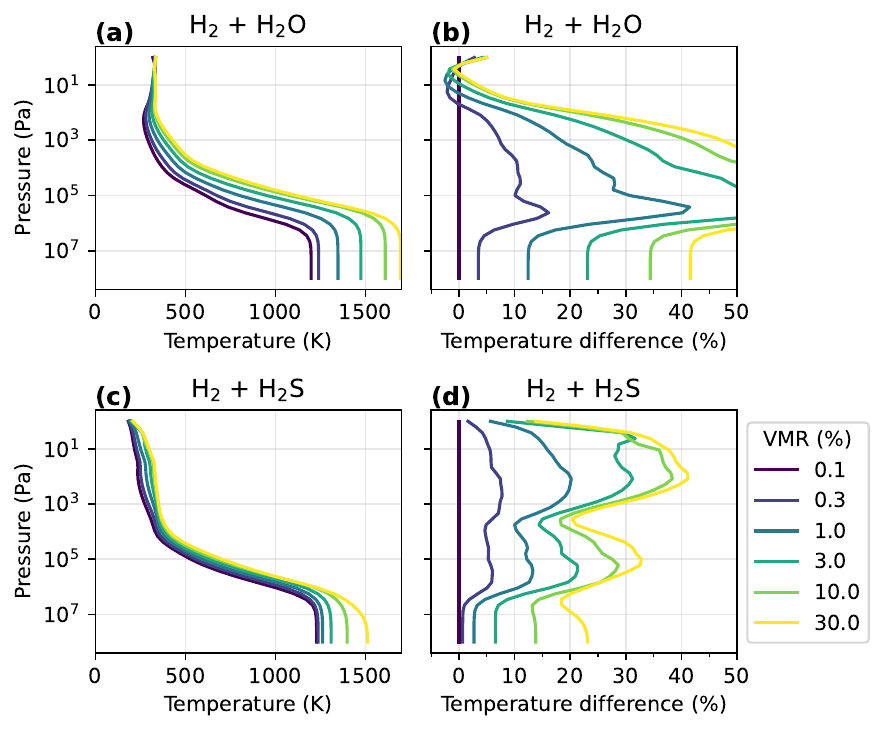}%
    \vspace*{1mm}
    \includegraphics[width=0.9\linewidth]{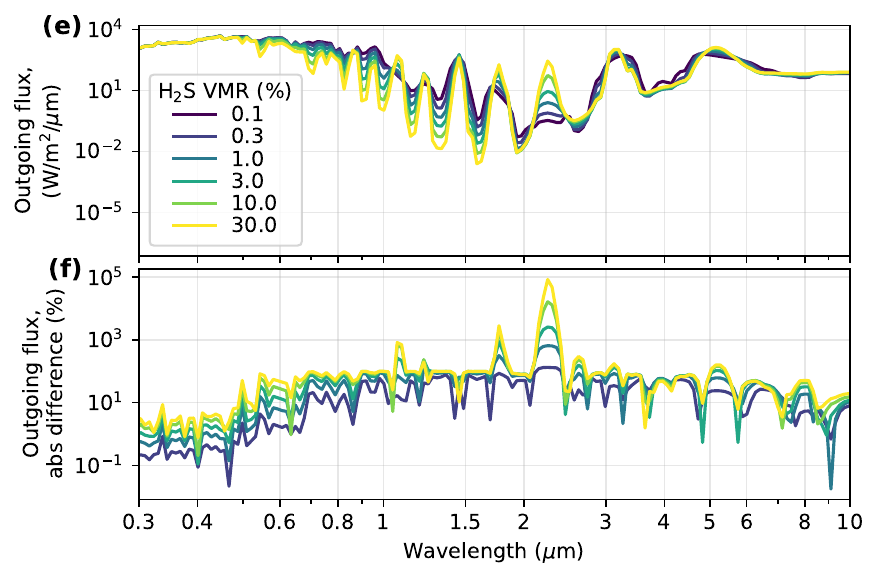}%
    \caption{Differences in atmospheric temperature structure and outgoing emission introduced by addition of H$_2$O or H$_2$S into a hydrogen background. \textbf{Top}: atmosphere absolute-temperature structures and relative differences for water-enhanced (a,b) and sulfur-enhanced (c,d) scenarios. \textbf{Bottom}: outgoing emission spectra arising from the sulfur-enhanced scenarios, and relative differences. Line colours show additive-gas volume mixing ratios. }
    \label{fig:sulfur}
\end{figure}

\section{Interior Parameters' Sensitivity Tests}
\label{apsec:generalsensitivitytests}

\textcolor{black}{Many of the parameters held constant in our study (table \ref{tab:constantparameters}) are necessarily assigned values by either adopting Earth-centric assumptions, or on an ad hoc basis, for lack of alternative constraints. We perform sensitivity tests for the subset of these parameters to which our results could be the most sensitive. Specifically, we perform tests where we vary the core density, core heat capacity, initial flux at the top of the mantle, initial specific entropy in the mantle, thickness of the conductive boundary layer, and the reference viscosity of the solid mantle. For these studies, all parameters that are not varied are held constant at the values specified in table \ref{tab:constantparameters}. The sensitivity of the model to these parameters is quantified using the melt mass-fraction of the upper regions of the mantle (R>0.5R$_{\text{int}}$) -- this property quantifies the melt content of a magma ocean which has solidified from the bottom-up. We now discuss the details of each sensitivity test and their results.}

\begin{figure}
    \includegraphics[width=\linewidth]{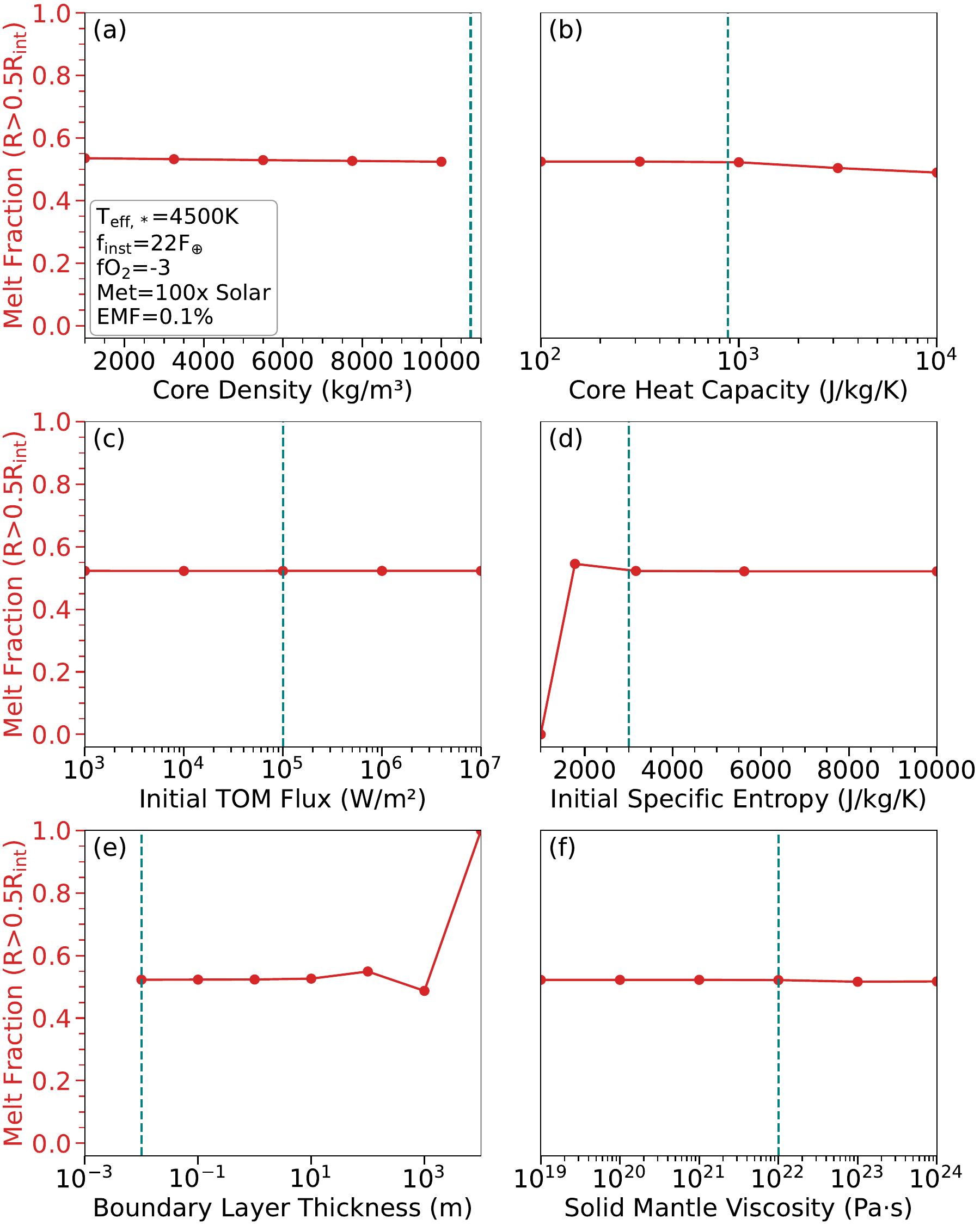}%
    \caption{The results of our sensitivity tests. The melt fraction above the middle mantle layer (R>0.5R$_{int}$) is shown as a function of core density (panel a), core heat capacity (panel b), initial top-of-mantle flux (panel c), initial specific entropy (panel d), thickness of the conductive boundary layer between the mantle and the surface (panel e) and the reference viscosity for the solid portion of the mantle (panel f). The values for each parameter used in the main study are shown in teal. All of the parameters that are not varied are held constant at their fiducial values specified in table \ref{tab:constantparameters}.}
    \label{fig:sensitivitytests}
\end{figure}




\textcolor{black}{\textit{Core Density}: Our models uses a core density taken from a preliminary reference earth model \citep{Bower2018}. However, sub-Neptunes are likely to have higher core densities than Earth, due to the extra compression resulting from higher planet masses. Additionally, if sub-Neptune core compositions were to deviate significantly from that of Earth's core, this would affect the core density \citep{Howe2014}. We vary the core density from 1000 kg m$^{-3}$, the density of water at room temperature and pressure, up to 11000\,kg m$^{-3}$. We choose 11000\,kg m$^{-3}$ as the upper limit in this test because densities larger than this result in deep mantle pressures too large for our numerical solver. The results of our test show that the near-surface melt fraction is insensitive to the density of the core (figure \ref{fig:sensitivitytests}). This is unsurprising, given that core density largely affects pressures in the deep mantle, which would only affect the deep-mantle melt content and the interior structure calculation, leaving the surface melt fraction (and our main discussion and conclusions) unaffected.}

\textcolor{black}{\textit{Core Heat Capacity}: Our model nominally adopts the 880 J kg$^{-1}$ K$^{-1}$ metallic-core specific heat capacity from \cite{Bower2018}, which represents an adjustment on the  preliminary reference earth model. The core heat capacity could vary for sub-Neptunes with core compositions that differ from that estimated for Earth; e.g. due to different melting phase-states or elemental compositions \citep{Howe2014}. We vary the modelled core heat capacity from 100 J kg$^{-1}$ K$^{-1}$ to 10000 J kg$^{-1}$ K$^{-1}$ to account for these unknowns. The results of this test show that the near-surface melt fraction is insensitive to the heat capacity of the core (figure \ref{fig:sensitivitytests}a). This is due to thermal equilibration between the core and the mantle over shorter timescales than the overall cooling timescale, leading to limited disequilibrium heat exchange between the core and the mantle.}

\textcolor{black}{\textit{Initial Top-of-Mantle Flux}: The flux upper-boundary condition for the mantle evolution model is a free parameter, which we set arbitrarily. However, this initial flux would depend on the formation history of the planet in question \citep{ikoma_constraints_2006, Lebrun2013}. The effect of this arbitrary initial condition on mantle evolution has not been investigated for sub-Neptunes. To test this, we vary the initial surface heat flux from 10$^{3}$ W m$^{-2}$ to 10$^{7}$ W m$^{-2}$, which correspond to fluxes calculated via the Stefan-Boltzmann law using surfaces temperatures of 364\,K and 3644\,K respectively. We find that the upper-mantle melt fraction is insensitive to the initial surface energy flux (figure \ref{fig:sensitivitytests}c). This is a consequence of our modelling approach. In subsequent time-steps of the modelled simulations, the flux upper boundary condition for the interior model is updated self-consistently by the energy conservation solver within our climate model \citep{Nicholls2025}. If the flux boundary condition on the upper-mantle at $t=0$\,yr deviates substantially from that which would be calculated by the climate code, the temperature structure of the mantle adjusts rapidly to compensate.}

\textcolor{black}{\textit{Initial Specific Entropy}: The initial specific entropy at the surface is a free parameter in our model, which we set arbitrarily. Previous work has shown that the evolution of Earth-like planets is insensitive to the initial specific entropy of the mantle \citep{Lichtenberg2021}. However, it remains unclear whether this result is also valid for sub-Neptunes, although they are still expected to form in substantially-molten states \citep{davies_silicate_2020, ikoma_constraints_2006}. We arbitrarily vary the mantle's initial value of specific entropy from 1000\,J K$^{-1}$ kg$^{-1}$ to 10000\,J K$^{-1}$ kg$^{-1}$. These simulations find that upper-mantle melt fraction is only sensitive to the initial entropy for low specific entropies <2000\,J K$^{-1}$ kg$^{-1}$ (figure \ref{fig:sensitivitytests}d). This is because an initial surface specific entropy less than 2000\,J K$^{-1}$ kg$^{-1}$ results in an initial mantle temperature structure below the solidus, i.e., the mantle is initialised solid. Therefore, provided the initial surface entropy is large enough to result in an initially molten mantle \citep{davies_silicate_2020,Lebrun2013}, the evolutionary outcome will not be affected by the initial surface entropy.}

\textcolor{black}{\textit{Boundary Layer Thickness}: The thickness of the boundary layer between the mantle and the surface is held constant in our model. However, previous theoretical studies have suggested that the thickness of a potential conductive surface-boundary layer would increase as the mantle solidifies \citep{Lebrun2013,schaefer_PREDICTIO_2016}. At the same time, other studies have discounted the presence of this layer entirely \citep{elkins_Linkedma_2008}. In principle, a conductive boundary layer could inhibit cooling of the mantle, given sufficiently large boundary layer thicknesses. We vary the thickness of the boundary layer from 0.01\,cm to 10\,km, given that theoretical modelling by \cite{Lebrun2013} showed that the thickness of this layer can increase to \textasciitilde\,km during late crystallisation. Our sensitivity study shows that the near-surface melt fraction can be sensitive to the boundary layer thickness when the thickness exceeds 1\,km (figure \ref{fig:sensitivitytests}e). This is due to an effective decoupling of the mantle thermal structure (and thus melt state) from the cooling at the surface, due to inefficient heat transport between the upper-mantle and lower-atmosphere. However, the presence of this theorised boundary layer is not guaranteed -- for example, it could be broken-up by atmosphere or interior dynamical shear stresses \citep{elkins_Linkedma_2008, Nicholls2024}. If sustained against dynamical factors, the boundary layer  thickness is only theorised to approach 10\,km during the later stages of mantle cooling \citep{Lebrun2013}. Therefore, our results are insensitive to the thickness of a potentially-present conductive boundary layer.}

\textcolor{black}{\textit{Reference Solid Viscosity}: In our calculation of the aggregate viscosity of the mantle \ref{equ:aggregateviscosity}, we use fixed reference solid and melt viscosities. However, the viscosity of Earth's solid mantle varies as a function of depth, temperature, grain size, water content, melt fraction, stress, and composition \citep{Cathles2015}. There is also uncertainty around the viscosity of molten peridotite \citep{DINGWELL2004}. Given that the cooling of an initially molten mantle is primarily regulated by the opacity and blanketing effect of an outgassed atmosphere, rather than the ability of the molten mantle to transport heat to the surface \citep{Lebrun2013,schaefer_PREDICTIO_2016,Schaefer2017}, empirical uncertainties in melt viscosity are unlikely to affect our results. However, once the mantle begins to solidify, the solid viscosity begins to affect heat transport in the mantle. Therefore, uncertainties in the solid viscosity could affect our results. To test this, we vary the reference solid viscosity between 10$^{19}$ and 10$^{24}$ Pa s, to reflect the range of viscosity values in Earth's mantle \citep{Cathles2015}. We find that the near-surface melt fraction is insensitive to the reference solid viscosity (figure \ref{fig:sensitivitytests}f). This is a consequence of the fact that, for these sub-Neptunes, mantle fractionally solidifies from the bottom-up, with complete solidification occurring at the base of the mantle first \citep{Tang2024}. Therefore, the melt state of the mantle near the surface is unlikely to be affected by the viscosity of solid material at depth.}

\bsp	
\label{lastpage}
\end{document}